\documentclass[sigconf]{acmart}
\usepackage{popets}

\setcopyright{popets}
\copyrightyear{YYYY}

\acmYear{YYYY}
\acmVolume{YYYY}
\acmNumber{X}
\acmDOI{XXXXXXX.XXXXXXX}
\acmISBN{}
\acmConference{Proceedings on Privacy Enhancing Technologies}
\usepackage{tikz}
\usepackage{amsmath}

\newcommand \offline[0]{static}
\newcommand \realtime[0]{dynamic}
\newcommand \offlineXserver[0]{$static^+$}
\newcommand \realtimeXserver[0]{$dynamic^+$}
\newcommand \BI [0]{LR}
\newcommand \BII[0]{IR}
\newcommand \BIII[0]{FP}
\newcommand \BIV[0]{MU}

\usepackage{pifont}
\usepackage{graphicx}
\usepackage{subcaption}
\usepackage{multicol}
\usepackage{algorithm}
\usepackage{algpseudocode}
\usepackage{hyperref}
\usepackage{float}
\usepackage{stfloats} 
\usepackage{amsmath}
\usepackage{multirow}
\usepackage{pgf}
\usepackage{tikz}
\usepackage{svg}
\usepackage{xcolor}
\usepackage{amsthm} 

\newtheorem{hypothesis}{Hypothesis}
\newtheorem{definition}{Definition}
\DeclareMathOperator*{\argmax}{arg\,max}
\def\BibTeX{{\rm B\kern-.05em{\sc i\kern-.025em b}\kern-.08em
    T\kern-.1667em\lower.7ex\hbox{E}\kern-.125emX}}

\begin{document}

\title{Gradient-Free Privacy Leakage in Federated Language Models through Selective Weight Tampering}

\author{Md Rafi Ur Rashid}
\orcid{0000-0002-3089-5277}
\affiliation{%
  \institution{Pennsylvania State University}
  \country{}
}
\email{rafiurrashid@psu.edu}

\author{Vishnu Asutosh Dasu}
\affiliation{%
  \institution{Pennsylvania State University}
  \country{}}
\email{vdasu@psu.edu}

\author{Kang Gu}
\affiliation{%
  \institution{Dartmouth College}
  \country{}}
\email{f003hy4@dartmouth.edu}

\author{Najrin Sultana}
\affiliation{%
  \institution{Pennsylvania State University}
  \country{}}
\email{nks5814@psu.edu}

\author{Shagufta Mehnaz}
\affiliation{%
  \institution{Pennsylvania State University}
  \country{}}
\email{smehnaz@psu.edu}

\renewcommand{\shortauthors}{Trovato et al.}

\begin{abstract}
    Federated learning (FL) has become a key component in various language modeling applications such as machine translation, next-word prediction, and medical record analysis. These applications are trained on datasets from many FL participants that often include privacy-sensitive data, such as healthcare records, phone/credit card numbers, login credentials, etc.  
    Although FL enables computation without necessitating clients to share their raw data, existing works show that privacy leakage is still probable in federated language models. 
    In this paper, we present two novel findings on the leakage of privacy-sensitive user data from federated large language models without requiring access to gradients. Firstly, we make a key observation that model snapshots from the intermediate rounds in FL can cause greater privacy leakage than the final trained model. Secondly, we identify that a malicious FL participant can aggravate the leakage by tampering with the model's selective weights that are responsible for memorizing the sensitive training data of some other clients, even without any cooperation from the server. Our best-performing method\footnote{Our source code: \href{https://figshare.com/s/8636bd193495193c0fe4}{https://figshare.com/s/8636bd193495193c0fe4}} increases the membership inference recall by 29\% and achieves up to 71\% private data reconstruction, evidently outperforming existing attacks that consider much stronger adversary capabilities. Lastly, we recommend a balanced suite of techniques for an FL client to defend against such privacy risk.
\end{abstract}


\maketitle

\section{Introduction}  
\label{intro}
Large language models (LLMs) are playing a pivotal role in various fundamental natural language processing (NLP) tasks~\cite{joshi2017triviaqa, kieuvongngam2020automatic}. Recently, these LLMs have been trained on extremely large text
corpora~\cite{gao2021pile} and grown to billions of parameters \cite{brown2020language,touvron2023llama}. These pre-trained LLMs are then adapted to
various downstream tasks by fine-tuning on domain-specific datasets, often on privacy-sensitive datasets such as in-house emails \cite{huang2022large}, code repositories \cite{bigcode}, clinical health data \cite{jagannatha2021membership}, etc. Given the privacy-sensitive nature of these datasets, collecting them by a centralized entity to fine-tune an LLM raises significant privacy concerns. To solve such issues, federated learning \cite{shokri2015privacy, sui2020feded, tian2022fedbert} has emerged as a widely-used technology. FL allows language models to be trained by exchanging only model updates, without requiring a central entity to collect all the training data.
However, recent studies \cite{wang2019beyond,  zhao2020idlg, wen2025sok} have shown that carefully crafted attacks can invert the model updates sent by FL clients and recover private information. To make matters worse, LLMs tend to output complete sequences verbatim from their training data~\cite{carlini2021extracting, carlini2023quantifying}, which by itself is a major privacy concern. 

Most prior work investigating the privacy of language models focuses on centralized settings \cite{cheng2025effective, carlini2021extracting,carlini2019secret,gu2023towards} that often do not transfer to FL due to the fundamental differences in their training protocols. For instance, in a centralized setting, the training process \emph{sees} all the data points in every epoch, whereas FL uses asynchronous aggregation \cite{nishio2019client}, where the server selects a subset of clients, each with a fraction of the data, in each round of training. 
Besides, most existing research on FL privacy leakage is either strictly applicable for finite class-based downstream tasks~\cite{hitaj2017deep}, assumes the server as the adversary instead of FL clients~\cite{zhao2024loki}, or only infers certain properties from training data (e.g., person wearing glasses, gender, age)~\cite{melis2019exploiting}. They neither transfer to generative AI tasks, such as language modeling, nor do they extract private training sequences verbatim, as a malicious FL client. 
There are a few works that focus on privacy attacks in federated language models~\cite{vu2024analysis}, but they assume the role of an untrusted server to extract general training data from FL users~\cite{fowl2022decepticons, gupta2022recovering} and require access to gradients~\cite{lamp}, rather than devising more impactful attacks as a malicious client that specialize in extracting privacy-sensitive data, e.g., PII (Personally Identifiable Information).
To fill this research gap, in this paper, we investigate the leakage propensity of users' \textbf{privacy-sensitive} data that a malicious client is particularly interested in.  
For instance, email service providers (e.g., Gmail, Outlook) may train an FL model for their auto-complete feature that learns from the users' emails without requiring direct access to email data, thus providing the privacy promised by FL. However, users may share various types of privacy-sensitive data, such as SSNs or credit card numbers, in their emails. It would be a serious privacy concern if one user of such a federated setting could extract sensitive information about other users. Our findings show that this is indeed possible, where the attacker is curious about some specific type of sensitive data to extract from other users. Such extracted data could then be used to undertake serious security and privacy crimes, e.g., abusing extracted credit card information, performing SSN reverse lookup, and then impersonating the victim to banks/other institutions.

In most FL algorithms, such as FedSGD \cite{mcmahan2016communication}, model gradients are shared with a central server in clear text, thereby creating a potential attack vector for privacy breaches. In such settings, adversaries can readily exploit gradient information, and numerous gradient-leakage attacks \cite{du2024sok, leakage_gradients, huang2021evaluating, jiang2022comprehensive} have emerged as state-of-the-art techniques for red-teaming machine learning models. 
In this work, we deal with a much tighter FL implementation, i.e., federated averaging (FedAvg), where the gradient information is kept private and only weights are shared \cite{mcmahan2016communication}. Previous research on information leakage \cite{hatamizadeh2023gradient} and inversion attacks \cite{huang2021evaluating, elkordy2022much} in FL concludes that FedAvg is significantly harder to attack or leaks less privacy than FedSGD. Also, due to the high communication overhead of FedSGD, FedAvg and its variants—such as FedProx, FedAdam, and FedYogi—are more commonly adopted in practice~\cite{li2020federated, reddi2020adaptive}. 
Nevertheless, understanding the extent of privacy leakage in FL-based language models is complex and challenging. This is because previous research on forgetting analysis \cite{jagielski2022measuring} shows that deep language models tend to gradually forget early training samples if they are not repeated much afterward. Additionally, as seen in studies on catastrophic forgetting \cite{french1999catastrophic, kemker2018measuring}, during iterative training, models can lose information about earlier samples as they encounter new ones. Since only a subset of the clients participates in each round of FL, the above observations are inherently true for FL settings. The FedAvg algorithm also reduces the memorization as the training heads toward convergence \cite{thakkar2020understanding}.
 These observations imply that the model tends to forget any client's local data as training proceeds and tries to generalize more on the overall distribution of the entire training corpus. 
This could potentially make it difficult for an attacker to extract a client's private data from the \textbf{final model} in FL. \emph{However, existing research has not yet explored the privacy vulnerabilities related to the \textbf{intermediate model} snapshots in federated language models.}
Hence, in our attack design, we exploit the model snapshots from intermediate rounds of FL, where the participating clients have privacy-sensitive text, e.g., phone numbers, SSNs, passwords, etc. in their local datasets. We show how a malicious FL participant can retrieve some other FL participants' sensitive data even without the server's cooperation by maximizing the memorization of those intermediate model snapshots.

Besides exploiting the FL configuration's intrinsic vulnerability in our attack design, we further investigate how an attacker could enhance the model's memorization of privacy-sensitive training data. {The proposed methods stem from the crucial observation that \emph{some weights of an LLM are more susceptible to sensitive data points than others due to their \textbf{out-of-distribution} nature.} Specifically, the MLP sub-layers in the last few transformer blocks have the strongest influence \cite{mireshghallah2022memorization, chen2024learnable}. Therefore, strategically adjusting these selective weights in an LLM can effectively improve its ability to memorize privacy-sensitive data. 
Based on this observation, we designed two methods for tuning the model weights and piloting them toward stronger memorization. The first one is a \emph{zeroth-order} method called \textbf{selective weights optimization (SWO)}, which takes into account the difference of selective weights between a fine-tuned (with privacy-sensitive data) model and another reference model, and then maximizes this difference. The second approach, named \textbf{weights transformation learning (WTL)}, utilizes an end-to-end supervised algorithm to learn the hidden pattern of how those selective weights mutate from a reference model to a fine-tuned model. The attacker may apply these techniques either actively during fine-tuning (referred to as \emph{dynamic} attack) or passively after the fine-tuning is done (referred to as \emph{static} attack). Our \emph{dynamic} mode of the attack closely resembles the recent line of works using model poisoning for Privacy Leakage \cite{zhang2023agrevader, 10190537, fowl2021robbing, wen2024privacy, feng2024privacy}. 
However, most of these existing works consider only centralized learning scenarios and poison the pre-trained model beforehand \cite{wen2024privacy, feng2024privacy, liu2024precurious}. They also presume strong adversarial capabilities, e.g., assuming the server to be an \emph{active} adversary  \cite{nguyen2023active}, accessing the gradients \cite{10190537}, or changing the model's architecture \cite{fowl2021robbing}. In contrast, our \emph{static} attack, with no model poisoning or server cooperation, results in up to 63\% sensitive data reconstruction, while our \emph{dynamic} attack, involving active weight tampering during FL fine-tuning, improves the reconstruction rate up to 71\%.
Both attack modes are highly effective on open-source LLMs, including Gemma \cite{team2024gemma}, Llama-2 \cite{touvron2023llama}, GPT-2 \cite{radford2019language}, and BERT \cite{devlin-etal-2019-bert}.
We also emphasize that our attacks are not domain-specific. They are applicable as long as the sensitive texts are somewhat \emph{out-of-distribution} in nature, such as any domain-specific jargon, medical prescriptions, and forensic test results. 
Although previous works on Privacy Leakage attacks exploited the outlier nature of private texts \cite{carlini2019secret, huang2022large, lukas2023analyzing}, ours is the first work to maximize data memorization of a federated language model by tampering with its selective weights to increase the privacy leakage. Finally, we experiment with possible privacy defenses, including differential-privacy \cite{dwork_roth_dp} , regularization \cite{wu2023depn}, and scrubbing \cite{chowdhury2021adversarial}, and recommend the best possible combination of methods for an FL client to defend against such privacy leakage.
Our main contributions are as follows:

\ding{114} We propose a gradient-free novel attack framework that exploits model snapshots from intermediate rounds in FL to leak the privacy-sensitive text data of the FL users.


\ding{114} We introduce two novel methods- SWO and WTL to maximize LLM's memorization by tampering with selective weights, which are largely responsible for memorizing privacy-sensitive data.


\ding{114} Our proposed framework functions as a \emph{pre-attack routine}, bolstering both membership inference and reconstruction attacks on clients' private data, significantly outperforming existing attacks.

\ding{114} We empirically analyze possible defenses and recommend the best suite of methods to mitigate such privacy threats.
\section{Background}
\textbf{Data Reconstruction Attack:} It is defined on the verbatim extraction of a subsequence from the model’s training dataset. For the autoregressive text generation, we consider every sensitive sequence $x = p||c$ in the dataset, where $x$ is split into a prefix $p$ (i.e., the non-sensitive component) and a suffix $c$ (i.e., the sensitive component). The suffix is the attacker's point of interest.
For each sequence, if the model exactly reproduces $c$ when prompted with $p$, it is considered a successful reconstruction. In masked language modeling, for every sensitive sequence $x = p||c||s$ in the training dataset, the attacker's aim is to reconstruct the masked-out sensitive component $c$, given a prefix $p$ and suffix $s$.

\noindent \textbf{Membership Inference:} In this attack, the attacker attempts to infer whether a sensitive sequence $x$ is part of the victim's local training data or not. In this work, we evaluate the sentence-level membership inference introduced by Mireshghallah et al. \cite{mireshghallah2022quantifying}.

\noindent \textbf{Perplexity:} Perplexity~\cite{gonen2023demystifying} is a widely used metric to measure how well the LM predicts tokens in the generated sentence. For a sequence of predicted tokens $x_1, \ldots, x_n$, perplexity is computed as:
{\footnotesize
\[
    \mathcal{P} = \exp \left(-\frac{1}{n}\sum_{i=1}^{n}\log \theta(x_i | x_1, \ldots, x_{i-1})\right) \notag
\]}
where $\theta$ is the language model. A low perplexity implies that the model is confident, i.e., less surprised by the generated text.

\noindent \textbf{Exposure:} Carlini et al.~\cite{carlini2019secret} first introduced exposure as a quantitative measure of how much a generative LM has unintentionally memorized a rare training sequence/canary. Concretely, they (i) compute a sequence’s log-perplexity under the model, (ii) define the \emph{rank} of an inserted canary, and (iii) set exposure to the negative log-rank, which can be interpreted as the reduction in guessing difficulty an attacker gains from access to the model. Higher exposure, therefore, means the canary has become disproportionately likely relative to other equally formatted candidates, signaling greater risk of targeted extraction. We use this metric, and defer full mathematical details to Appendix \ref{exp_res}. Additional background concepts, including FL, autoregressive and masked language models, and differential privacy, are provided in Appendix \ref{app:bg}. 

\section{Related Work}

\subsection{Privacy Leakage Attacks in LLMs}
Privacy risks of LLMs have been extensively studied in prior works. The seminal work by Carlini et al. \cite{carlini2021extracting} proposes an attack to retrieve public internet texts by querying a pre-trained GPT-2 model. 
Our attack, however, aims to extract privacy-sensitive data when fine-tuning an LLM in FL. In the domain of RNN and LSTM models, 
Carlini et al. \cite{carlini2019secret} highlight the problem of unintended memorization. They insert canaries into the training dataset and measure their leakage probability with a metric called `exposure', which we also used in our work to evaluate our novel attack strategies. Mireshghallah et al. \cite{mireshghallah2022memorization} study the effect of fine-tuning on memorization and show that fine-tuning the parameters higher in the LLM architecture results in the most data leakage. Our experiments in FL align with this observation, and we manipulate these selective weights that are most susceptible to private data for producing greater memorization.  
Cheng et al.~\cite{cheng2025effective} pre-conditioned the LLM with structured PII examples and used an online learning loop to improve extraction success. They defined in-training PII as findable on today’s web, which is not always applicable, because some outputs may be simple inferences or guesses based on public patterns.
Akkus et al.~\cite{akkus2025generated} showed that fine-tuning LLMs using generated data without an instructional structure can amplify PII leakage from their pre-training dataset, while
Kim et al. \cite{kim2023propile} and Lukas et al. \cite{lukas2023analyzing} investigated to what extent information memorized by the LM constitutes sensitive PII and whether existing defenses are sufficient to prevent leakage. Although their objectives are similar to ours in this work, they differ in two crucial aspects: (1) they are conducted under a \emph{non-FL, centralized} threat model, where the central server is considered the adversary, whereas we focus on privacy leakage caused by the malicious participants in FL, (2) most of these works are evaluation based, measuring the extent of privacy leakage caused by finetuning/pretraining LLMs, while we propose dedicated methods to amplify such leakage besides its evaluation.

\subsection{Privacy Leakage By Poisoning}
Recent research has explored model and data poisoning to leak privacy, though most studies do not focus on language modeling \cite{zhang2023agrevader, 10190537}. Rashid et al.~\cite{rashid2025forget}, Wen et al. \cite{wen2024privacy}, Feng et al. \cite{feng2024privacy}, and Liu et al. \cite{liu2024precurious} manipulate the weights of pre-trained LLMs to cause privacy leakage during fine-tuning. However, these attacks do not apply to FL. On the other hand, Fowl et al. \cite{fowl2021robbing} consider the server in FL as an adversary that actively poisons model architecture and parameters sent to users. Such attacks are easily detectable in pre-trained LLMs, where the architecture is well-known to clients. In contrast, our \emph{static} attack method can passively leak the privacy of FL users after fine-tuning without poisoning the pre-trained model, thus fully preserving the utility of the model. Additionally, our \emph{dynamic} attack further enhances Privacy Leakage by actively tampering with the model's weights during FL training, where the attacker is one of the FL clients, not the server.

\subsection{Privacy Leakage in Federated Learning}
Among existing works on FL privacy leakage~\cite{du2024sok, hitaj2017deep, vu2024analysis}, many have attempted to invert gradients to extract training data~\cite{du2024sok, nasr2019comprehensive}. Recent Gradient Inversion Attacks (GIAs), such as Gifd~\cite{fang2023gifd}, Gi-pip~\cite{sun2024gi} and Spear~\cite{dimitrov2024spear} reconstruct image/vision inputs by optimizing dummy data so that their induced gradients match the victim’s shared gradients. Due to the non-differentiable nature of discrete text, the GIA threat to language models (LMs) is more limited than for images, but Li et al.~\cite{li2025temporal} reconstructs private text by exploiting gradients from token-embedding layers, while the LAMP \cite{lamp} attack inverts the client's gradients into natural text by alternating between continuous and discrete optimization. Critically, most GIAs assume server access to per-client gradients, which aligns naturally with \textbf{FedSGD}~\cite{mcmahan2016communication}. In contrast, our work uses a more standard and commonly used FL algorithm--\textbf{FedAvg}~\cite{li2020federated, reddi2020adaptive}, which shares only the final local model after multiple minibatch steps, so the server at best sees an aggregated, multi-step parameter delta rather than the exact client gradients--making classic GIAs much less applicable in practical FedAvg deployments.
Apart from that, Hitaj et al.~\cite{hitaj2017deep} adversarially train a GAN to reconstruct class-representative images from the client's private data. However, it is strictly applicable for finite class-based downstream tasks, such as image classification, and does not transfer to generative tasks like language modeling. Besides, Melis et al.~\cite{melis2019exploiting} exploit gradients of the embedding (for text), convolutional or fully-connected layer (image) to infer certain properties from client's training data, such as gender, age, someone's facial attributes in an image etc, but unlike our attacks, they solely focus on classification task, and do not extract victim's private texts verbatim. Hu et al.~\cite{hu2025simple} sample PII-contextual and frequency-prioritized prefixes to extract sensitive client data. Although they follow a similar probing strategy to ours, their methods exploit only the model's inherent memorization of PII rather than amplifying it.
Decepticons \cite{fowl2022decepticons}, Loki~\cite{zhao2024loki} and Vu et al.~\cite{vu2024analysis} are three attacks that assume a malicious FL server with the
ability to manipulate the model architecture and parameters
before sending the model to clients, while FILM~\cite{gupta2022recovering} considers an external adversary that eavesdrops on the communication channel between the server and clients to reconstruct sentences from gradients. Our attack differs from theirs in that we neither treat the server as an active adversary nor assume any side-channel information leak.} Section~\ref{benchmark} presents a detailed comparison of our attack with Decepticons and FILM. 

\section{Impact of Privacy-Sensitive Data on Fine-tuning Large Language Models}
\label{impact}

\begin{figure}[t]
\centering
    \begin{subfigure}[b]{0.5\linewidth}
    \includegraphics[width=\linewidth]{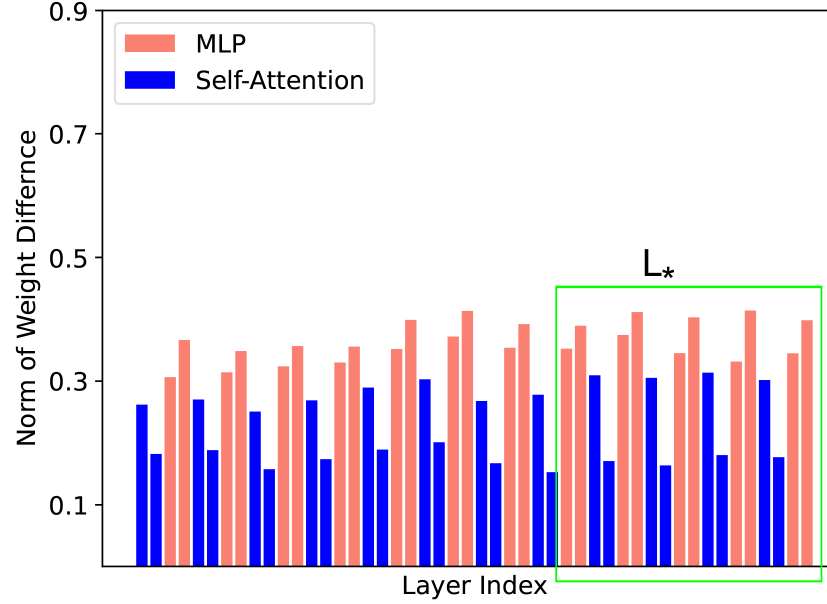}
    \caption{}
    \label{fig:weights_diff_gen}
    \end{subfigure}
    \begin{subfigure}[b]{0.48\linewidth}
    \includegraphics[width=\linewidth]{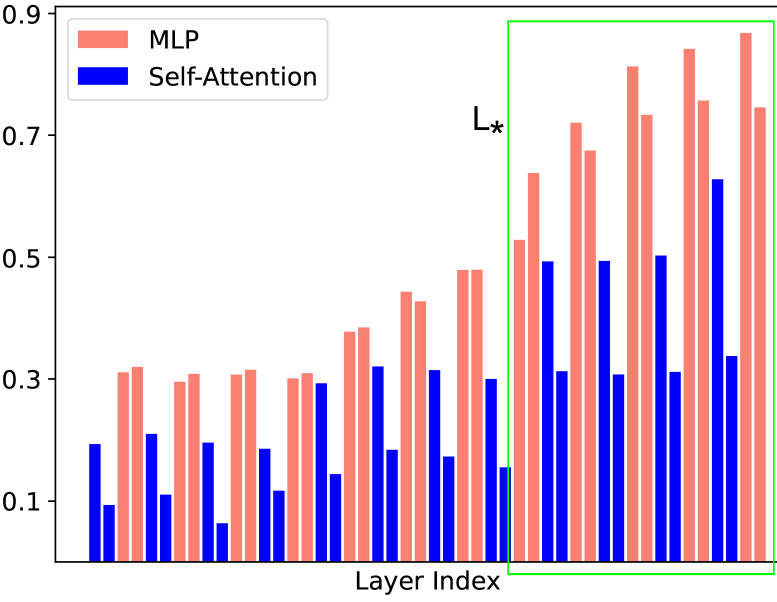}
    \caption{}
    \label{fig:weights_diff_sens}
    \end{subfigure}
    \caption{Norms of  weight changes in  MLP and self-attention layers of all 12 transformer blocks after fine-tuning GPT-2 with (a) regular English texts and (b) privacy-sensitive texts}
\end{figure}

To carry out different downstream applications, LLMs such as GPT-2 \cite{radford2019language} and BERT \cite{devlin-etal-2019-bert} undergo a fine-tuning step that modifies their pre-trained weights according to those specific tasks \cite{chung2022scaling}. 
However, privacy-sensitive texts have specific patterns and often contain out-of-distribution tokens, e.g., digits (in phone or credit card numbers), special characters, and out-of-vocabulary words (login credentials or physical addresses). 
To understand the impact of such privacy-sensitive data on fine-tuning, we investigate whether the changes in the pre-trained weights after fine-tuning the LLMs with such sensitive data show any differentiating patterns when compared to the weight changes after fine-tuning with regular English texts. Our key observation from this investigation is that there indeed exists a remarkable difference between the patterns in these two cases. 
Figure \ref{fig:weights_diff_gen} depicts the norms of weight changes in all 12 transformer blocks, i.e., multi-layer perceptron (MLP) and self-attention layers after fine-tuning a GPT-2 (base) model $G$ with regular English sentences $\mathcal D_{reg}$. Here, we observe a uniform impact on the weights across different layers, $L$, which suggests that the model's adaptation after fine-tuning is evenly distributed throughout the architecture. 
However, a distinct pattern emerges when fine-tuning $G$ with privacy-sensitive texts $\mathcal D_{sen}$ such as phone numbers, email addresses, login credentials, credit card numbers, medical prescriptions, etc. Only a subset of weights, specifically the MLP sub-layers in the last few transformer blocks, denoted as $L_{\star} \subset L$, undergo a substantial transformation, while the remaining layers, denoted as $L_{\ominus}= L\setminus L_{\star}$, exhibit minimal/regular changes. This exact observation is also supported by Chen et al. \cite{chen2024learnable}. Figure \ref{fig:weights_diff_sens} shows the transformation of the MLP and self-attention layers of GPT-2 after fine-tuning with $D_{sen}$ (Refer to Figure \ref{fig: gemma_weight_diff} for Gemma).
We experiment with four different language models - Gemma, Llama-2, GPT-2, and BERT - using various types of sensitive data and find that as long as these sensitive texts have an \textbf{out-of-distribution} characteristic with respect to the training data, each of these LMs shows such disproportionate changes in their internal weights. This crucial observation supports the following hypothesis:

\begin{hypothesis} 
\label{hyp:weights_change} 
 Weights in some selective layers of an LLM undergo significant changes compared to the other layers while fine-tuned on out-of-distribution data.
\end{hypothesis}

Formally, after fine-tuning with privacy-sensitive data, if $\Delta W_{\star}$ and $\Delta W_{\ominus}$ represent the mean weight changes in respective layers, they can be defined as below:
\begin{center}
\footnotesize
$\Delta W_{\star} = \frac{1}{|L_{\star}|} \sum_{l\in L_{\star}} (W(f_G(\mathcal D_{sen}))-W(f_G(\cdot))) | _l$\\
$\Delta W_{\ominus} = \frac{1}{|L_{\ominus}|}\sum_{l\in L_{\ominus}}(W(f_G(\mathcal D_{sen}))-W(f_G(\cdot))) | _l$\\ 
\end{center}
Here, $W(f_G(D))$ denotes the weights of model $G$ fine-tuned with dataset $D$.
We observe that $\|\Delta W_{\star}\| >> \|\Delta W_{\ominus}\|$. In contrast, when fine-tuning with regular texts, we observe $\|\Delta W_{\star}\| \sim \|\Delta W_{\ominus}\|$.
This observation indicates that weights of certain layers ($L_{\star}$ marked with green rectangles in Figures \ref{fig:weights_diff_gen} and \ref{fig:weights_diff_sens}) within an LLM are more susceptible to capturing and memorizing privacy-sensitive information. 
We also scrutinize the direction/pattern of changes in those selective weights caused by out-of-distribution samples in the fine-tuning dataset. Then we ask the most important question- \emph{if we further increase the amplitude of those particular weight changes along the same direction, does it make the model more biased towards the out-of-distribution data?} Later, in Section \ref{max_exp} where we discuss our Maximizing Data Memorization (MDM) techniques, we illustrate that the answer to this question is `Yes', and we can strategically focus on modifying the weights of these sensitive layers to enhance the model's memorization of private data. But that is not all. With this also comes the salient question- \emph{If we amplify the weight changes in such a manner, does that affect the model utility?} We also convincingly answer this question in Section \ref{max_exp} by showing that such modification hardly impacts the model's learning of regular English texts.
In summary, the adversary aims to find a function $\mathcal{T}: W_\star \rightarrow W'_\star$ that generates a new set of weights $W'_\star$ which will maximize the memorization of $\mathcal{D}_{\text{sen}}$. 
\begin{equation}
\label{eq:5}
\footnotesize
\underset{\mathcal{T}}{\argmax}~\mathbf{M}(\mathcal{D}_{\text{sen}},   \mathcal{T} (W_{\star}))
\end{equation}
We qualitatively define memorization $\mathbf{M}$ as a property of the LM in exposing particular training data points during inference. Quantitative measure of memorization $\mathbf{M}$ on $\mathcal{D_{\text{sen}}}$ includes but are not limited to data reconstruction \cite{carlini2019secret}, membership inference \cite{mattern-etal-2023-membership}, and exposure, which we discuss in detail in Section \ref{measure_leakage}.

\section{Attack Methodology}
\label{s:methodology}
\subsection{Threat model}
\label{threat}
In our settings, both the adversary and the victim(s) are members of the client group participating in FL to fine-tune the LLM for a text auto-completion task, where they don't share any gradient information with each other.
We assume that the adversary has some context of the victim's training data, specifically, the non-sensitive prefixes of the privacy-sensitive texts they want to extract from the victim's local dataset. This assumption is highly practical in corporate and administrative contexts where communication relies on boilerplate templates. For example, automated notification emails ('Your verification code is...'), medical intake forms, or banking alerts follow rigid structural patterns. Similar to the probing strategy in several existing attacks~\cite{kim2023propile, liu2024precurious, carlini2021extracting, hu2025simple}, here the adversary only needs to know the standard template (prefix) to extract the unique sensitive suffix. Section \ref{sec:impl} discusses more on our template-based prompt design. 
Apart from that, we consider two independent cases of the adversary's capability: (1) the adversary is a stand-alone malicious client in the FL setting who does not collude with the server and observe the global model in each round, and (2) optionally, the malicious client can collaborate with the server which enables it to exploit a specific victim's local model. 

\begin{figure*}[t]
    \centering
    \includegraphics[width=0.8\textwidth]{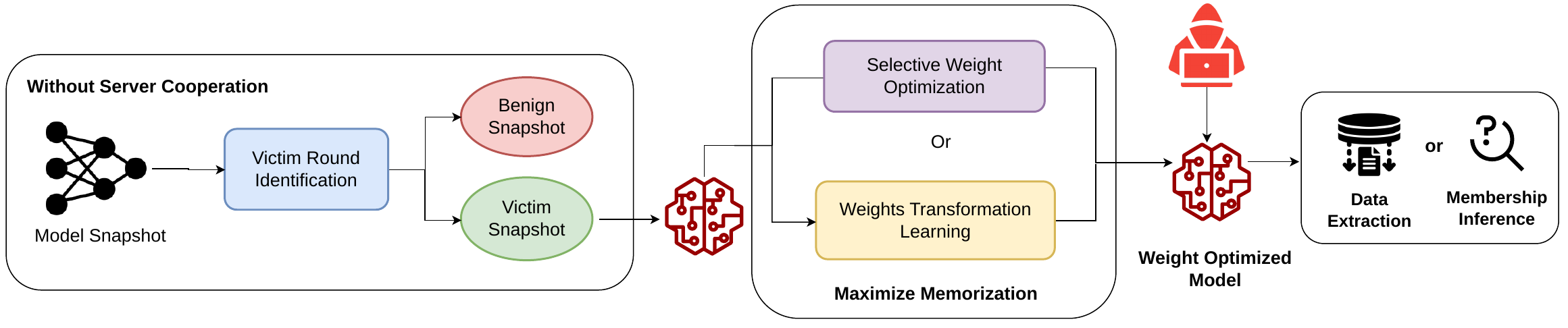}
 \caption{
 Overview of our attack flow. First, we identify the training rounds where the victim participated using the \emph{VRI} algorithm. Next, we update the victim models to maximize the memorization of sensitive information using \emph{MDM} algorithms. Finally, we prompt the model with crafted prefixes to retrieve sensitive information.
 }
  \label{fig:attac_diagram}
\end{figure*}

Without server cooperation, the adversary extracts certain types of sensitive data without targeting a specific client, and consequently, all the users having such types of data in their local datasets fall victim to this attack. 
In contrast, with server cooperation, they can also target a particular victim and compromise their privacy. However, unlike many existing works that consider the server as an active adversary \cite{ nguyen2023active, fowl2022decepticons, fowl2021robbing}, we assume a more practical attack surface where the server is \textbf{passive} and helps the malicious participants by only sharing some meta information. This is a more practical setting since the server has some accountability, and it could cause a loss of its reputation if it runs active attacks during the FL training.
Table \ref{tab:taxonomy} shows a high-level taxonomy of our attack methods. We delve into all the mentioned attack attributes in the later parts of this section. Figure~\ref{fig:attac_diagram} presents a high-level overview of our attack methods.
There are two vital steps in our general attack pipeline:
(1) \emph{Victim round identification (VRI)}: The adversary identifies the FL training rounds where victims with privacy-sensitive text of the adversary's interest participated and
(2) \emph{Maximizing data memorization (MDM)}: After the FL training is complete, the adversary manipulates the identified victim rounds' models to maximize the memorization of sensitive data.

\begin{table}[t]
\centering
\resizebox{\linewidth}{!}
{
\begin{tabular}{c|c|c|c|c|c|c}
\hline
\textbf{\begin{tabular}[c]{@{}c@{}}Attack  \\ Attributes\end{tabular}} & \begin{tabular}[c]{@{}c@{}}\textbf{Server}\\ \textbf{Colllusion}\end{tabular} & \begin{tabular}[c]{@{}c@{}}\textbf{Adversary's}\\ \textbf{Action}\end{tabular} & \begin{tabular}[c]{@{}c@{}}\textbf{MDM}\\ \textbf{Method}\end{tabular} & \textbf{Victim}     & \begin{tabular}[c]{@{}c@{}}\textbf{Type of}\\ \textbf{LM}\end{tabular}             & \textbf{Goal}                                                           \\ \hline
\multirow{2}{*}{\textbf{Variations}}                                   & No                                                       & Static                                                          & SWO                                                  & Targeted   & \begin{tabular}[c]{@{}c@{}}Autoregressive\\ (GPT-2)\end{tabular} & \begin{tabular}[c]{@{}c@{}}Data\\ Reconstruction\end{tabular}  \\ \cline{2-7} 
                                                                      & Yes                                                      & Dynamic                                                           & WTL                                                  & Untargeted & MLM (BERT)                                                       & \begin{tabular}[c]{@{}c@{}}Membership\\ Inference\end{tabular} \\ \hline
\end{tabular}
}
\caption{A high-level taxonomy of our attack methods}
\label{tab:taxonomy}
\end{table}
\subsection{Victim Round Identification (VRI)} 
\label{sec:vri}
\begin{figure*}[t]
    \begin{subfigure}{0.262\textwidth}
        \includegraphics[width=\textwidth]{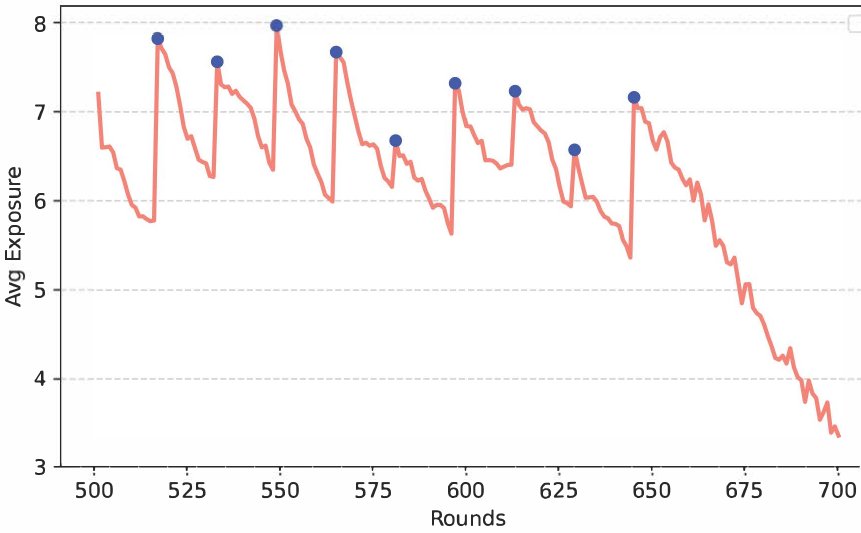}
        \caption{}
        \label{fig:exposure_trend}
    \end{subfigure}
    \hfill
    \begin{subfigure}{0.245\textwidth}
        \includegraphics[width=\textwidth]{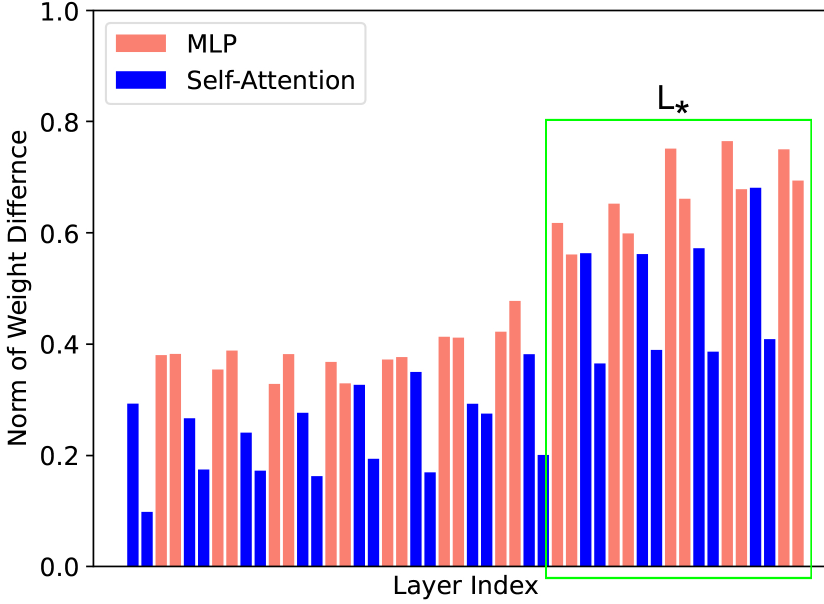}
        \caption{}
        \label{fig:weight_diff_vic}
    \end{subfigure}
    \hfill
    \begin{subfigure}{0.236\textwidth}
        \includegraphics[width=\textwidth]{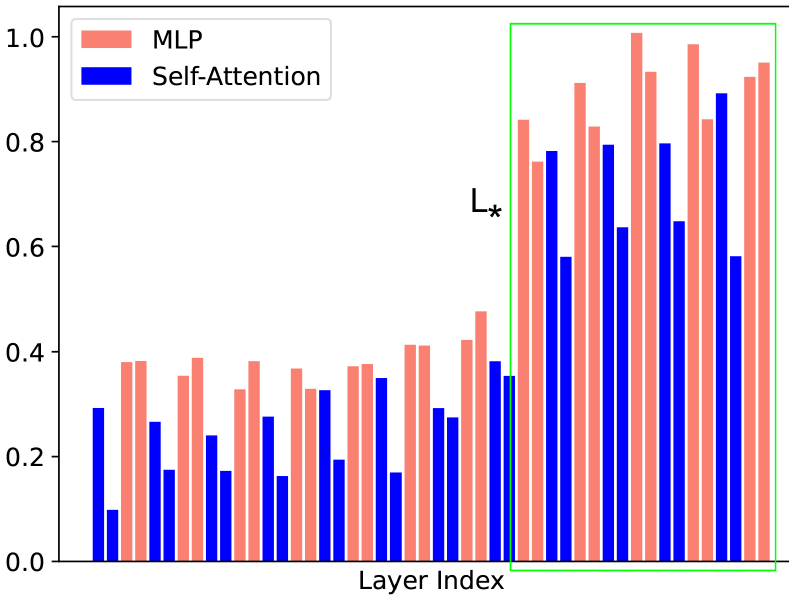}
        \caption{}
        \label{fig:weight_diff_vic_swo}
    \end{subfigure}
    \hfill
    \begin{subfigure}{0.236\textwidth}
        \includegraphics[width=\textwidth]{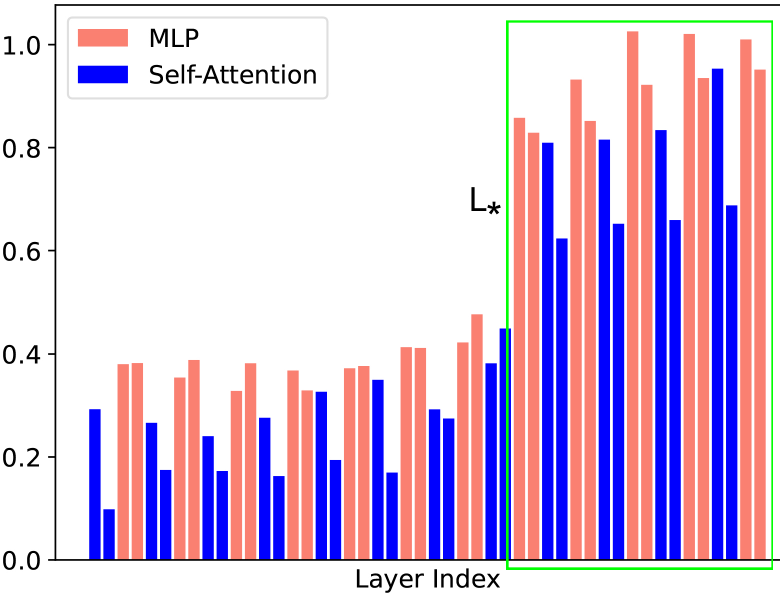}
        \caption{}
        \label{fig:weight_diff_vic_wtl}
    \end{subfigure}
    \caption{(a) Change of exposure of the private data with the passage of 200 FL rounds. The blue dots indicate those rounds when the victim participated. (b)-(d) Norms of the weight changes in the MLP and self-attention layers of all 12 transformer blocks of a victim snapshot, (b) without MDM, (c) with SWO, (d) with WTL}

\end{figure*}

As mentioned earlier, the attacker aims to extract some specific sensitive texts that follow certain patterns from the victims' local datasets. If multiple clients contain such private data of the attacker's interest, all of them fall victim to the attack. The objective of the VRI step is to identify some of the training rounds (not necessarily all) in which such victims participate.
As mentioned in Section \ref{intro}, LLMs tend to forget specific samples from an individual client's training data as they tend to generalize the overall data distribution during the training process.  To better realize this phenomenon, we exploit the exposure metric from \cite{carlini2019secret} and compute the average exposure (EXP) of the global model snapshots with regard to some victims' private data for all rounds of FL. As shown in Figure~\ref{fig:exposure_trend}, the value of EXP is higher for the rounds where the victim(s) participated (i.e., victim rounds, marked with blue dots) compared to the other rounds where they did not.  Furthermore, EXP gradually decreases between two victim rounds, and more obviously, there is a persistent fall in the value of EXP after round 650 since the server did not select any victim in any round afterward (or any victim may not have participated). Based on our observation of this significant difference between the EXP values in victim and non-victim rounds, we want to deduce the following hypothesis:

\begin{hypothesis}
    \label{hyp:victim_rounds} 
    The likelihood of a federated language model memorizing a sensitive data point in a training round is much higher if the data point is included in that round's training data compared to rounds where the data point is absent.
\end{hypothesis}

Hence the identification of global model snapshots from victim rounds can add a new dimension to the attacks focusing on the victim's privacy leakage in FL. To exploit this key observation in our proposed attacks, when we assume the server does not cooperate with the adversary, we devise a mechanism for victim-round identification (VRI).

The key idea behind our VRI algorithm is to differentiate between the victim and typical (non-victim) rounds' snapshots based on the characteristics of $L_{\star}$.
Let $G_i$ be the global model snapshot at round $i$.
We mark it as a victim snapshot $\widehat{G_i}$ if we identify that the victim participated in round $i$.

\underline{\textbf{Step 1:}}
The adversary stores the global model snapshots, $G_i$, where $i$ represents any of the $N$ training rounds when the attacker participates.

\underline{\textbf{Step 2:}} 
We fine-tune a pre-trained LLM instance $G$ with $\mathcal{D}_{reg}$ and denote it as $G_r$. Next, we fine-tune $G$ with $\mathcal{D}_{reg} \cup \mathcal{D}_{sen}$, i.e., regular English and sensitive texts combined, and denote it as $G_s$. 

\underline{\textbf{Step 3:}}
We calculate the norm of weight differences for each of the $L_{\star}$ layers ( based on hypothesis \ref{hyp:weights_change}) between $G$ and $G_r$, denoted as $\delta_r$, and between $G$ and $G_s$, denoted as $\delta_s$ as given below: 
\begin{center}
\footnotesize
$\delta_r = (\|W(f_{G_r}(\mathcal D_{reg}))- W(f_G(\cdot))\|)| _{L_{\star}}$\\
$\delta_s = (\|W(f_{G_s}( \mathcal D_{reg} \cup \mathcal D_{sen}))- W(f_G(\cdot))\|)| _{L_{\star}}$
\end{center}
We similarly calculate the norm of weight differences for $L_{\star}$ between $G$ and each $G_i$ and denote it as $\delta_i$. Our goal is to identify whether $G_i$ is a victim round.
\begin{center}
\footnotesize
$\delta_i = (\|W(f_{G_i}(\cdot))- W(f_G(\cdot))\|)| _{L_{\star}}$
\end{center}
Here, each of $\delta_r$, $\delta_s$ and $\delta_i$ is a vector of size $L_{\star}$.

\underline{\textbf{Step 4:}}
Based on hypothesis \ref{hyp:victim_rounds}, our key intuition of this final step is that if $G_i$ is a victim round snapshot, $\delta_s$ and $\delta_i$ would have a similar pattern in their distribution due to the presence of sensitive sequences in both of their training data, whereas $\delta_r$ and $\delta_i$ would have a significantly different distribution. For a better understanding, refer to Figures \ref{fig:weights_diff_gen} and \ref{fig:weights_diff_sens}.
Hence, in this step, we perform two pairwise t-tests between $\delta_i$ and $\delta_r$, and between $\delta_i$ and $\delta_s$, respectively. For a predefined significance level, if the test with $\delta_r$ rejects the null hypothesis ($H_0$) and the test with $\delta_s$ accepts it, we mark $G_i$ as a victim snapshot. Otherwise, $G_i$ is categorized as a non-victim snapshot. To verify that such an approach works, we found the T-test between $\delta_s$ and $\| \Delta W_{\star} \|$ of any victim model almost always (95\% cases) derive a p-value greater than the significance level, failing to reject the null hypothesis, which suggests an equivalent relationship between these two models. A detailed evaluation of our VRI method is provided in Section \ref{sec: vri}.
It is important to note that $\mathcal{D}_{sen}$ has no overlap with clients' private data. They are just some dummy out-of-distribution texts that have a similar pattern as the sensitive texts. 
More details on the curation of $\mathcal{D}_{sen}$ and $\mathcal{D}_{reg}$ are provided in the Attack Settings of Section \ref{sec:impl}.
We formally present our VRI technique in the Appendix (Algorithm \ref{alg:victim identify}). 

\subsection{Maximizing Data Memorization}
\label{max_exp}
After we identify victim snapshots ($\widehat{G_i}$), the next step is to maximize the victim models' memorization of the private data based on hypothesis \ref{hyp:weights_change} by transforming their selected weights ($W_{\star}$) as shown in Equation \ref{eq:5}. We design two approaches to achieve this: 

\begin{figure}[t]
    \begin{subfigure}[b]{0.5\linewidth}
    \includegraphics[width=\linewidth]{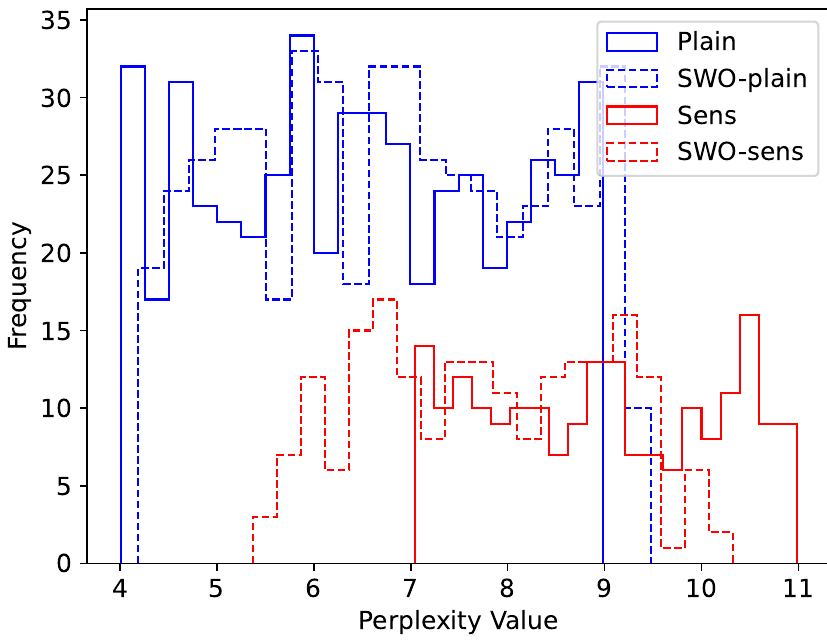}
    \caption{}
    \label{fig:swo_pplx}
    \end{subfigure}
    \begin{subfigure}[b]{0.48\linewidth}
    \includegraphics[width=\linewidth]{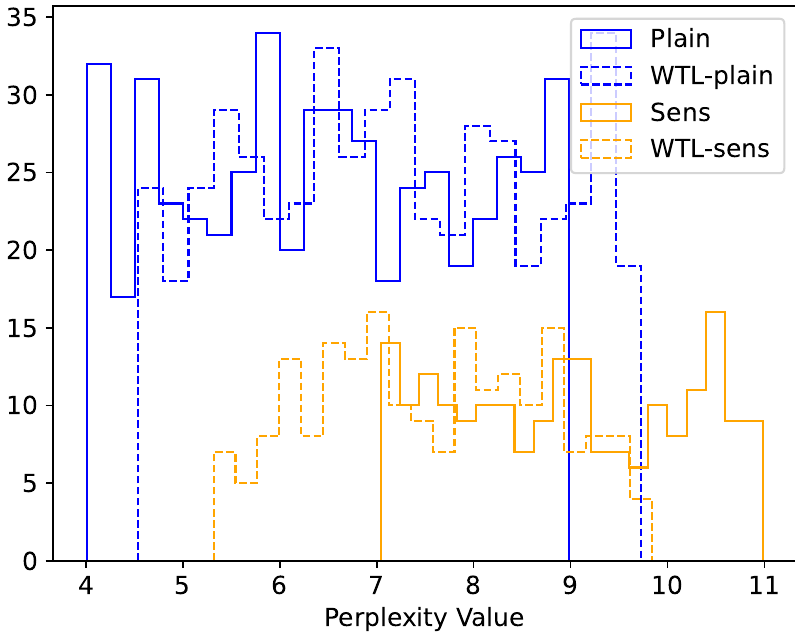}
    \caption{}
    \label{fig:wtl_pplx}
    \end{subfigure}
    \caption{Impact of MDM methods on data memorization and model's utility: (a) for SWO (b) for WTL. These results are generated by finetuning a GPT-2 base model with 1000 plain English texts and 200 out-of-distribution sensitive texts.}
    \label{mdm_pplx}
\end{figure}

\begin{algorithm}[]
\scriptsize
\caption{Selective Weights Optimization (SWO)}
\label{alg: weights_opt}
\begin{algorithmic}[1]
\Procedure{$\mathcal{J}$}{$W_1, W_2, \Delta W_{init}$}
    \State $\Delta W_{new} \gets W_1 - W_2 $
        \State \textbf{return} \hspace{1pt} $ \Delta W_{init} \cdot \Delta W_{new}$
\EndProcedure

\Procedure{Optimize\_Weights}{$W_1, W_2$, $\widehat{O}$}
    
    \State $\Delta W_{init} \gets W_1 - W_2$
    
    \For{$\text{step} := 1$ \textbf{to} $\text{num\_steps}$}
        \State $\mathcal{L}= -\mathcal{J}(W_1, W_2, \Delta W_{init}) $
        \State Backpropagate($\mathcal{L}$)
        \State $W_1 \gets \widehat{O}(W_1)$
    \EndFor
    
    \State \textbf{return} $W_1$
\EndProcedure

\For{each $W_1 \in W^{\widehat{G_i}}_{\star}$ \textbf{AND} $W_2 \in W^{G_r}_{\star}$}
    \State Initialize an Optimizer, $\widehat{O}$ with $W_1$ and a learning rate
    \State $W_1 \gets \textsc{Optimize\_Weights}(W_1, W_2, \widehat{O})$
\EndFor

\end{algorithmic}
\end{algorithm}
\noindent\ding{111} \textbf{Maximizing data memorization with \emph{Selective Weights Optimization (SWO)}:} We saw in the VRI step, for a victim round global snapshot $\widehat{G_i}$,  the t-test between $\delta_i$ and $\delta_r$ rejects the null hypothesis. Hence we hypothesize that there must be a distinction in $W_{\star}$ between $\widehat{G_i}$ and $G_r$, denoted as $\delta_v$.
{\footnotesize
\[
\delta_v = (\|W(f_{\widehat{G_i}}(\cdot))- W(f_{G_r}(\mathcal D_{reg}))\|)| _{L_{\star}}=\| W_{\star}^{\widehat{G_i}}- W_{\star}^{G_r}\|
\]}
This distinction in weight change characterizes how the model snapshot $\widehat{G_i}$ memorizes the privacy-sensitive sequences in contrast to the regular data. 
A higher difference essentially indicates higher memorization, resulting in more privacy leakage.   
Hence, we design an optimization problem that intends to maximize this difference ($\delta_v$) by updating $W_{\star}$ of $\widehat{G_i}$ ($W_{\star}^{\widehat{G_i}}$) to increase memorization of $\widehat{G_i}$ on sensitive data.  
Besides, the update is done only in magnitude, keeping the initial direction (+ve and -ve sign of the values in $W_{\star}$) intact so the model does not deviate from its origin arbitrarily. Algorithm \ref{alg: weights_opt} is a streamlined demonstration of the fundamental steps of our proposed optimization scheme. We define a custom objective function $\mathcal{J}(\cdot)$ that takes a dot product between $\Delta W_{init}$ and  $\Delta W_{new}$. Here, $\Delta W_{init}$ is the initial weight difference, and $\Delta W_{new}$ is the updated weight difference after each optimization step between $W_1 \in W^{\widehat{G_i}}_{\star}$ and $W_2 \in W^{G_r}_{\star}$. 
Later, this dot product is used as the cost measure $\mathcal{L}$ in the optimization phase, based on which we run the backpropagation. Here, the dot product ensures that updates are aligned with the direction of the initial state ($\Delta W_{init}$) and thus preserving the polarity of $W^{\widehat{G_i}}_{\star}$.

The optimization process can be performed using any conventional optimizer (e.g., SGD, Adam). The selective weights $W_1 \in W^{\widehat{G_i}}_{\star}$ are updated iteratively over a predefined number of steps. The optimizer is reset at each step, and the new difference ($\Delta W_{new}$) as well as the cost ($\mathcal{L}$) are calculated with the updated $W_1$. We take the negative cost measure in line 8 since we aim to maximize it. 
Additionally, the optimization may include
backpropagation with gradient clipping \cite{pascanu2013difficulty} and early stopping \cite{prechelt2002early} to prevent exploding gradients \cite{pascanu2013difficulty} and running extra steps during the update.
In Figure \ref{fig:weight_diff_vic} and \ref{fig:weight_diff_vic_swo}, we show the contrast before and after applying SWO on a victim snapshot ($\widehat{G_i}$), respectively. It is evident that the norm of $\Delta W_{\star}$ has increased. 
Also Figure \ref{fig:swo_pplx} indicates that the perplexity scores of the sensitive texts significantly decrease after applying SWO, whereas the scores remain almost as before in the case of plain English texts.
Hence, we can deduce that with the selectively optimized weights, $\widehat{G_i}$ becomes more vulnerable to privacy leakage of the sensitive sequences without hurting its utility. Experimental results in Section \ref{results} also advocate the above phenomenon.
\begin{algorithm}[]
\scriptsize
\caption{Weights Transformation Learning (WTL)}
\label{alg: weights_trans}
\begin{algorithmic}[1]
\Procedure{Learn\_Transform}{$G$, $\mathcal{D}_{reg}$, $\mathcal{D}_{sen}$, $n$}
\State $W_r \gets \{\}$
\State $W_s \gets \{\}$
\For{$i:= 1$ to $n$}
    \State $G_{r}^i \leftarrow$Fine-tuned $G$ with $\mathcal{D}_i\in \mathcal D_{reg}$
    \State $G_{s}^i \leftarrow$ Fine-tuned $G$ with $\mathcal{D}_i\in \mathcal D_{reg} \cup D_{sen}$ 
    \State $ W_{\star}^{G_{r}^i} \gets W(f_{G_{r}^i}(\mathcal{D}_i\in \mathcal D_{reg})) | _{L_{\star}}$
    \State $ W_{\star}^{G_{s}^i} \gets W(f_{G_{s}^i}(\mathcal{D}_i\in \mathcal D_{reg} \cup \mathcal D_{sen})) | _{L_{\star}}$
    \State $W_r \gets W_r \cup W_{\star}^{G_{r}^i}$ 
    \State $W_s \gets W_s \cup W_{\star}^{G_{s}^i}$ 
    
\EndFor
\State Initialize a supervised model $\mathcal{M}$
\State Train $\mathcal{M}$ with $W_r$ as input and $W_s$ as output
\State \textbf{return} $\mathcal{M}$
\EndProcedure
\State $\mathcal{M} \gets \textsc{Learn\_Transform}(G, \mathcal{D}_{reg}, \mathcal{D}_{sen}, n)$
\For{each $W \in W^{\widehat{G_i}}_{\star}$}
    \State $W \gets \mathcal{M}(W)$
\EndFor
\end{algorithmic}
\end{algorithm} 

\noindent\ding{111} \textbf{Maximizing data memorization with Weight Transformation Learning (WTL):} If we have a closer look at Figures \ref{fig:weight_diff_vic} and \ref{fig:weight_diff_vic_swo}, we observe that although the magnitude of impact on $L_{\star}$ is magnified after applying SWO, it does not fully coincide with the expected trend. In other words, the fluctuations of the magnitudes of $\| \Delta W_{\star} \|$ in Figure \ref{fig:weight_diff_vic_swo}  differ from that of Figure \ref{fig:weight_diff_vic}. A possible reason behind this could be modifying \emph{all} the weight values of $W_{\star}$ as described above during our optimization process for maximizing the difference. To verify this hypothesis and also to get a fine-grained understanding of how the underlying weight values transform while fine-tuned with sensitive data, we do further analysis and come up with weight transformation learning (WTL) using an end-to-end supervised algorithm. The high-level idea is to train a supervised model to capture the hidden pattern of how the weights in $L_{\star}$ change in the presence of sensitive training data.

Algorithm \ref{alg: weights_trans} gives a simplified demonstration of this transformation learning technique. Firstly, we need to generate $n$ (number of samples to train the supervised model $\mathcal{M}$) copies of $G_r$ and $G_s$ by fine-tuning $G$ with different sets of $\mathcal{D}_{reg}$ and $\mathcal{D}_{sen}$, respectively. Here, $D_{sen}$ does not need to be from the same distribution as the victim’s private data because according to hypothesis \ref{hyp:weights_change}, the weights in $L_{\star}$ undergo similar changes for any out-of-distribution data. 
We get $n$ samples of $W_{\star}$ from both $G_r$ and $G_s$, denoted as $W_r$ and $W_s$, respectively. Finally, we train the supervised model $\mathcal{M}$ with $W_r$ as input and $W_s$ as output.
Once $\mathcal{M}$ is trained, we can feed $W_{\star}$ of $\widehat{G_i}$ to $\mathcal{M}$ and get the updated weights. In Figure \ref{fig:weight_diff_vic} and \ref{fig:weight_diff_vic_wtl}, We show the contrast before and after applying WTL on a victim snapshot ($\widehat{G_i}$). Besides increasing the magnitude of impact for $L_{\star}$ compared to Figure \ref{fig:weight_diff_vic_swo}, it also shows a better alignment with the trend shown in Figure \ref{fig:weight_diff_vic}. Besides, in Figure \ref{fig:wtl_pplx}, we observe that the language model's memorization of the sensitive texts increases (perplexity goes down) by a higher margin with WTL when compared with SWO in Figure \ref{fig:swo_pplx} and the model's utility is also preserved.
\subsection{Optional: Attacks With Server Collusion}
\label{s:with_server}
The server can be \textit{honest-but-curious}, a threat model that has been widely investigated in prior works \cite{zhao2020idlg,wang_hbc_server,lamp}. In such cases, the server can passively cooperate with the adversary to enhance the victim participants' privacy leakage but does not actively run the attack to remain stealthy. The advantage of including the server under the adversary is twofold. Firstly, it is important to note that 
the victim/non-victim round information is readily available to the parameter server.
Although we have designed a victim-round identification mechanism as described earlier in Section~\ref{sec:vri}, it may incur some errors in practice (see Table \ref{tab:vic_idn}). In contrast, the server can provide the adversary with ground-truth victim round information-- thus making the privacy leakage attacks more successful. 
Secondly, in the FL setting, the adversary receives global model updates from the parameter server in every iteration. As discussed in Section \ref{max_exp}, the adversary aims to maximize the memorization of the victim's privacy-sensitive data in those global model updates. However, it can be easily inferred without loss of generality that the impact of the victim's data will be higher on its \textbf{local} model's weights compared to the \textbf{global} model's weights because local updates of all participants go through federated averaging, which weakens the victim's impact on the global model. In other words, direct access to the local models may provide a greater advantage to the adversary in leaking the victim's privacy-sensitive information. Therefore, the server can share the victim's local model updates with the malicious clients, who can then maximize its memorization of private data using the techniques described in Section \ref{max_exp}.

\subsection{Static and Dynamic Modes of Attacks}
We develop our attacks in two modes: \emph{\offline}~and \emph{\realtime}.
We denote the attacks with \emph{\offlineXserver}~and \emph{\realtimeXserver}, respectively, where the server cooperates with the adversary.

\noindent\textbf{Static Mode:} In this mode, the adversary does not actively interfere during the FL training but instead passively exploits the model after the fine-tuning is completed. It involves two actions: 

    \ding{111} After getting a global model update from the parameter server, it attempts to identify whether it is a victim round. With server cooperation, this step can be omitted, as the server shares the victim's local updates with the adversary.

    \ding{111} After the entire training is complete, the adversary attempts to maximize the memorization of all identified victim snapshots using one of the MDM techniques discussed in Section \ref{max_exp}. 

\noindent\textbf{Dynamic Mode}
Here, the adversary maliciously alters its local updates to magnify memorization of victims' private data during FL training.
It involves the following two actions:
    
    \ding{111} The adversary aggregates all the victim's snapshots they have identified so far, creating an aggregated model $G_A$.
    
    \ding{111} When he participates in a FL round, it maximizes memorization of $G_A$ using one of the two MDM methods. Later, instead of sending only its local update $G_L$ to the server, it sends a weighted aggregation of  $G_L$ and  $G_A$ as :
        $G_{LA} = \alpha G_A + (1-\alpha) G_L$
        
\noindent The weight $\alpha$ determines the contribution of $G_A$ to the final shared local update $G_{LA}$. For instance, if the adversary wants to send only $G_A$ without sharing their local updates, they set $\alpha$ to 1. 
However, with server cooperation, all of the victim's global updates are simply replaced with their local instances in this approach.    
Such active interference by the adversary has the potential to increase the overall attack performance as it infects the global model by pushing the victim's snapshot more frequently in the training process.

\subsection{Evaluating Privacy Leakage}
\label{measure_leakage}
\textbf{Data Reconstruction:} After the VRI and MDM steps, our data reconstruction attack extracts private data from the victim models by prompting with the associated context using beam search \cite{wiseman2016sequence}.
Since our primary objective is to generate text with zero diversity and maximum resemblance to the training data, greedy decoding may seem to be a preferred approach. However, experiments show greedy decoding is subject to trivial memorization \cite{carlini2021extracting}, e.g., repeated tokens and substrings, which often do not result in useful generations. 
We empirically find that using the beam search method with a small beam size (2-5) is more effective in generating private texts within the training data. This particular generation scheme causes a significantly higher leakage of private texts than the others. We take top-3 accuracy, i.e., the three highest-scoring outputs produced by the beam search for each reconstruction are considered as the candidate data points leaked from the training data (i.e., total 200 canaries*3=600 reconstructions) for our evaluation and some sample reconstructions are provided in Table \ref{tab:canary_exmp}. 

\noindent\textbf{Membership Inference:} Besides data reconstruction, we also aim to evaluate whether our proposed strategies enhance the performance of \emph{membership inference} (MI) \cite{shokri2017membership} attacks. 
For this purpose, we adopt the reference-based attack proposed in \cite{mireshghallah2022quantifying}, which involves calculating the percentage of private training samples that are correctly classified as training members out of a pool of both training and validation samples. To determine if a sample was part of the training set, two models are used: a target model, which is one of the victim models in our case, and a reference model. By comparing the likelihood probabilities from both models, a ratio is calculated, and if this ratio is below a certain threshold, the sample is considered part of the training set. 
We put the implementation details in the Appendix (Section \ref{MI_impl}) due to space constraints.

\section{Experiment Setup}
\subsection{Dataset Description}
\label{sec:data}
Our experiments include both canaries and real-world sensitive data. To observe how dataset size affects attack performance, we selected two datasets: the Penn Treebank (PTB) dataset \cite{marcus1993building}, which is a small dataset consisting of only 5MB of text from financial news articles, and the Wikitext-103 (Wiki) dataset \cite{merity2016pointer}, which is a comparatively larger dataset consisting of 181MB of text.
Similar to \cite{carlini2019secret}, we augment both of these datasets with out-of-distribution artificial canaries, which we manually crafted instead of web scraping to ensure that the pre-trained LLMs have not seen these texts before in their pre-training phase. These canaries contain dummy privacy-sensitive information. We insert four types of private information as canaries: phone number, credit card number, email, and physical address. Some examples of the canaries are provided in the Appendix (Table \ref{tab:canary_exmp}). It is worth mentioning that the training dataset may contain out-of-distribution texts that are not privacy-sensitive, such as product keys and hotline numbers. However, in our attacks, the adversary specifically designs the prompts to extract only the privacy-sensitive information they are interested in. The presence of other non-privacy-sensitive outlier texts in the dataset does not affect the effectiveness of the attack.

To confirm that our proposed method does not merely work for artificially inserted canaries, we also include the Enron email dataset \cite{klimt2004introducing} in our experiments, which has several naturally occurring sensitive pieces of information. It
is a well-known collection of emails that were exchanged among employees of the Enron Corporation, a U.S. energy company. It contains private information, e.g., phone numbers, addresses, login credentials, and other sensitive details about individuals. We only retain the body of the emails and pre-process them by filtering out all attachments and auto-generated emails. We select a subset of the resulting dataset (412MB) and manually identify 125 privacy-sensitive sequences. 
Several existing works \cite{carlini2019secret, lukas2023analyzing, gupta2022recovering} on LLM's privacy conducted their experiments with these three datasets.
For the Enron email dataset, we observe that the majority of the 125 private sequences have multiple occurrences. Approximately 29\% of them occur twice, 18\% occur three times, and 7\% occur four or more times. This implies that although state-of-the-art defenses like differential privacy \cite{dwork_roth_dp} provide guarantees under the assumption that records are unlikely to be duplicated, this may not be true for real-life datasets.
To simulate this phenomenon in the PTB and Wikitext datasets, we generate 200 unique canaries and insert them into the datasets with $k$ repetitions. 
Specifically, we consider $k\in{\{1, 10, 50\}}$, meaning one canary per $\sim{\{46K, 4.6K, 920, 460\}}$ training sequences for PTB, and per $\sim{\{3.5M, 350K, 70K, 35K\}}$ training sequences for Wikitext. 

\begin{figure}[t]
    \centering
    \includegraphics[width=0.75\linewidth]{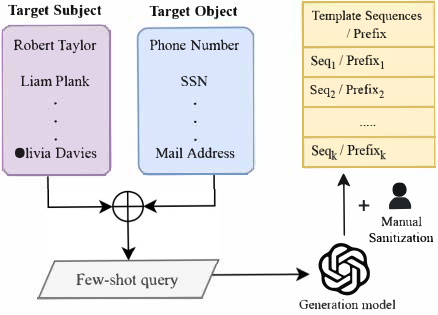}
 \caption{LLM-assisted construction of dummy sensitive sequences, $\mathcal{D}_{sen}$ and non-sensitive prefix templates for evaluating the data reconstruction attacks. }
 \label{fig:prompt_design}
\end{figure}
\subsection{Training and Attack Implementation} 
\label{sec:impl}
\noindent\textbf{FL Training Setup:}
We fine-tune Gemma-2b \cite{team2024gemma}, Llama-2 \cite{touvron2023llama}, GPT-2 base \cite{radford2019language} and BERT \cite{devlin-etal-2019-bert} as representative language models
to evaluate our proposed attacks. Later we also ablate for the medium and large versions of GPT-2. Due to computation resource limitations, we could not train the Llama-2 model in a large-scale FL setting. Instead, we conducted our attacks on Llama-2 in a compact experimental setup. Relevant experiments with Llama-2 are included in Appendix \ref{sec:llama-2}. Additional information about these models is available in the Appendix \ref{models}. To make fine-tuning large LMs feasible in FL, we adopt Parameter-Efficient Fine-Tuning (PEFT) \cite{hu2022lora} for all models exceeding 1B parameters. Specifically, for Gemma-2b and Llama-2, we use QLoRA configurations consistent with recent large-scale FL and PEFT benchmarks. We apply LoRA to the attention projection matrices and MLP components. Unless otherwise specified, we use LoRA rank $r=16$, scaling factor $\alpha=32$, and dropout $p=0.05$, with trainable parameters restricted to LoRA adapters while keeping the base weights frozen. Following Dettmers et al. \cite{dettmers2023qlora}, we quantize the base model to 4-bit NF4.
For the primary FL task, we have 100 participants and randomly select 20 for each round. We vary the number of attackers and victims between 1 and 10 and distribute the sensitive sequences among the victims accordingly. We train 
the model for 700 iterations, each with five local epochs, using an initial learning rate of $10^{-4}$ with a linearly decayed learning rate scheduler. 

\noindent\textbf{Attack Settings:} Details of attack settings are given below:

    \ding{111} We use a dataset of English plaintext jokes \cite{pungas} as $\mathcal{D}_{reg}$ as discussed in Sections \ref{impact} and \ref{s:methodology}. This dataset exclusively contains $1622$ regular English sentences without privacy-sensitive sentences. Besides, we generate $100$ dummy privacy-sensitive sentences to use as $\mathcal D_{sen}.$ As mentioned earlier, it does not overlap with the original private texts. As shown in Figure \ref{fig:prompt_design}, the attacker utilizes their prior knowledge of the sensitive data pattern to craft few-shot queries for a text generation LLM and generate these dummy sequences that capture the model's behavior towards out-of-distribution data.
    Additionally, the number of epochs for fine-tuning $G_r$ and $G_s$ was empirically set to 20, with early stopping enabled. 

    \ding{111} As depicted in Figure \ref{fig:weights_diff_sens}, the size of $L_{\star}$ is important.
    We empirically choose the top 18 (for GPT-2, BERT), 20 (for Gemma), and 28 (for Llama-2) layers that show the highest norm as $L_{\star}$. Later, we also perform an ablation study by varying the size of $L_{\star}$.
    
    \ding{111}  We use the Adam \cite{kingma2014adam} optimizer in the selective weights optimization algorithm.
    We empirically tune the number of steps between 200 and 1000, the clipping threshold between 1 and 4, and set the initial learning rate at $1 \times 10^{-3}$.
    
    \ding{111} Since model weights do not fall into conventional categories of data, we experiment with several learning algorithms for the transformation task to identify the most suitable one. These algorithms include normalizing flows (e.g., planar flows \cite{Lai-planner-flow} and RealNVP\cite{dinh2016realNVP}), dense autoencoder (DAE), variational autoencoder (VAE)\cite{kingma2013VAE}, convolutional autoencoder (CAE)\cite{zhang2018CAE}, and transformer\cite{vaswani_etal}. We find that DAE, CAE, and transformer showed comparable performances (in terms of validation loss) while normalizing flows and VAE did not yield promising outcomes.
    However, we eventually chose to employ DAE for this task as the other two, particularly the transformer, turned out to be computationally much more expensive taking roughly twice the time of DAE to complete the training. 
    
    \ding{111} To ensure better stability during the transformation, we use Z-score normalization \cite{al2018effect} to normalize the weight vector before passing it to the SWO or WTL algorithm. After the transformation, we denormalize the output weight to its original scale. For the sake of simplicity, we skip the normalization step in Algorithms \ref{alg: weights_opt} and \ref{alg: weights_trans}.
     
    \ding{111} For data reconstruction, we follow a similar probing approach to many existing works~\cite{liu2024precurious, kim2023propile, carlini2021extracting} that utilize non-sensitive prefix templates to extract sensitive suffixes from private training data. Although most works constructed prefixes via manual or frequency-based sampling~\cite{hu2025simple, liu2024precurious}, we adopt an LLM-based generation pipeline followed by manual sanitization. As shown in Figure \ref{fig:prompt_design}, the attacker aims to extract certain objects, such as phone numbers and SSNs, that belong to targeted subjects. Given those target subjects and objects, we query the GPT-4o API model to generate $K$ different sequences with different combinations of target subjects and objects. To ensure format consistency (i.e., prefix $\oplus$ suffix template), we include few-shot examples into the query, e.g., ``The police should urgently reach out to <Subject's name> by calling at <dummy phone number>'' or ``<Subject name> has a savings account in Bank of America, associated with the SSN: <dummy SSN>''. Following the few-shot query, the model generates $K$ sequences, which undergo a second format check via manual validation. Finally, the prefix portions are separated from these sequences using regular expressions for probing purposes in our data reconstruction attack.

\subsection{Baselines and Metrics}

We introduce four baseline scenarios to conduct a comparative analysis of different attack strategies:

\ding{111} \textbf{Last Round (LR)}: This is a naive attack scenario where the adversary has access to only the final model achieved after completing FL training, with no additional attack strategies applied.

\ding{111} \textbf{Intermediate Rounds (IR)}: The attacker has access to the ground truth information of intermediate victim rounds from server cooperation as mentioned in Section \ref{s:with_server}. IR directly prompts these victim models without applying any attack strategy. 

\ding{111} \textbf{Forced Participation (FP)}: As one of the extreme scenarios, we assume that the server intentionally manipulates the FL protocol and client selection process. Instead of random selection, the server forces the victims to participate in each round. This unfair action gives the highest importance to the victims' local data.
Since the victims' local updates are getting exposed in each iteration, it is not necessary to consider the intermediate rounds. Instead, we consider the final model after the training is completed. 

\ding{111} \textbf{Malicious Update (MU)}: In the strongest attack case, the server not only manipulates the client selection, but it also manipulates what the victim receives from the server in each iteration. The server keeps returning the victim's local update from the previous iteration to it instead of the global update. This forces the victim to fall into a vicious cycle of overfitting that continues till the training ends. 
In this case, we consider the final local update of the victim for the same reason as FP.

The first scenario will help us understand why we chose to exploit the intermediate snapshots of victim rounds instead of the final model (hypothesis \ref{hyp:victim_rounds}). The second scenario is to imply how much leakage the intermediate victim rounds cause without applying our proposed attack methods. On the other hand, \BIII~  and \BIV, with strong assumptions (i.e., less practical) on the adversary's capability, set an upper bound to our attack performance. 

\textbf{Metrics}. We report the number of successful reconstructions (NSR) of the artificial canaries (for Wikitext and PTB) and the inherent sensitive texts (for Enron). We also evaluate the level of leakage using exposure (EXP), a metric introduced in \cite{carlini2019secret}, where higher exposure indicates greater privacy leakage. For membership inference, we report the recall and precision following \cite{mireshghallah2022quantifying}. Moreover, we use validation perplexity (PPX) as a metric for the performance of the model, where lower perplexity refers to better model utility.
\begin{figure*}[htbp]

    \begin{subfigure}{0.255\textwidth}
    \includegraphics[width=0.98\textwidth]{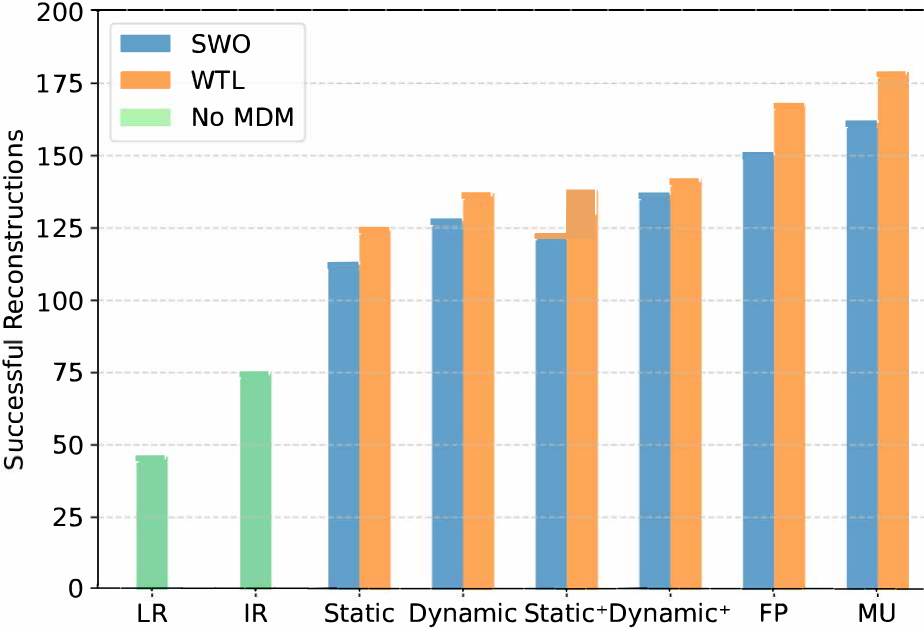}
    \caption{}
    \label{fig:Gemma_rec_accross_dft_level}
    \end{subfigure}%
    \begin{subfigure}{0.25\textwidth}
    \includegraphics[width=0.98\textwidth]{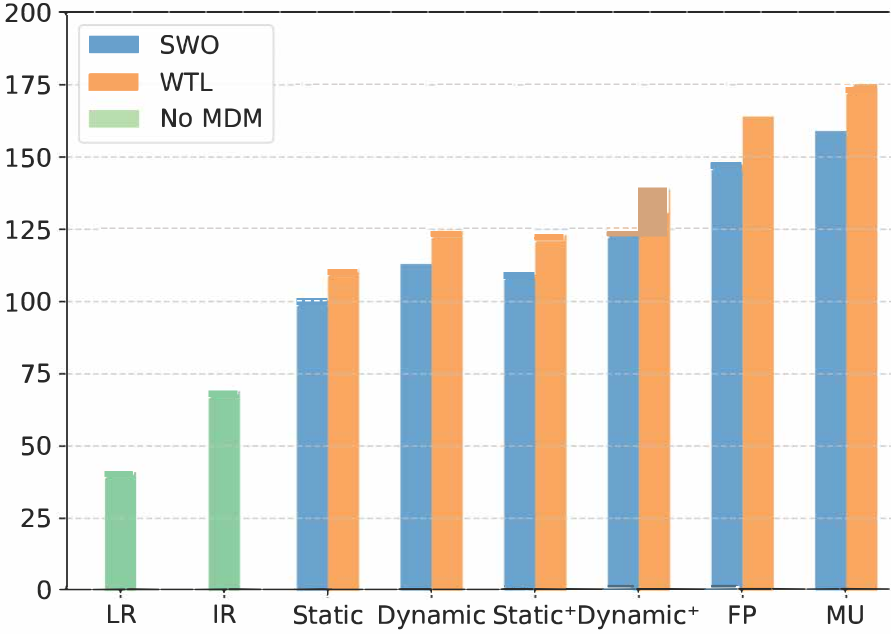}
    \caption{}
    \label{fig:rec_accross_dft_level}
    \end{subfigure}%
    \begin{subfigure}{0.25\textwidth}
    \includegraphics[width=0.98\textwidth]{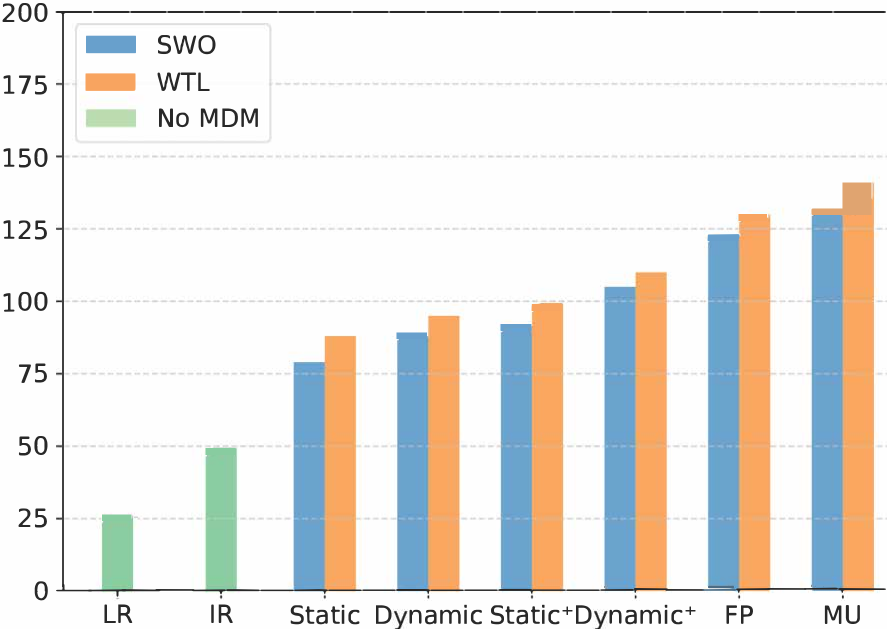}
    \caption{}
    \label{fig:mlm_rec_accross_dft_level}
    \end{subfigure}%
    \begin{subfigure}{0.25\textwidth}
    \includegraphics[width=0.98\textwidth]{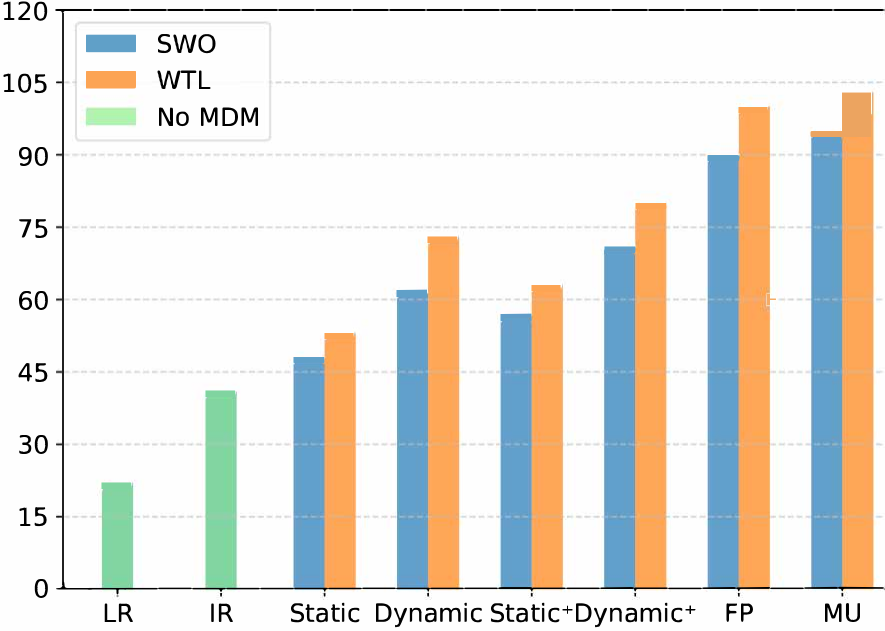}
    \caption{}
    \label{fig:enron_Gemma_rec_accross_dft_level}
    \end{subfigure}%

    \begin{subfigure}{0.25\textwidth}
    \includegraphics[width=0.98\textwidth]{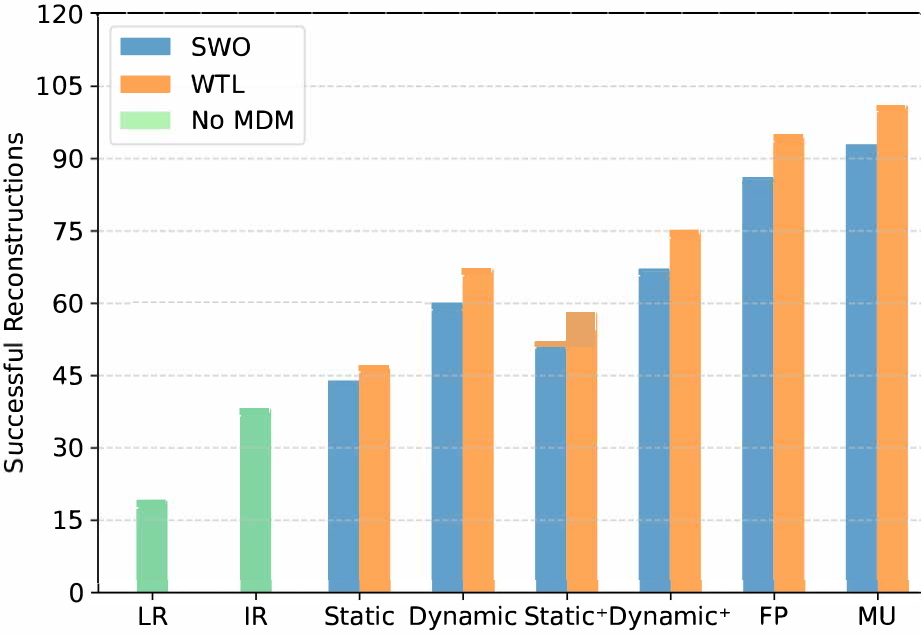}
    \caption{}
    \label{fig:enron_rec_accross_dft_level}
    \end{subfigure}%
    \begin{subfigure}{0.25\textwidth}
    \includegraphics[width=0.95\textwidth]{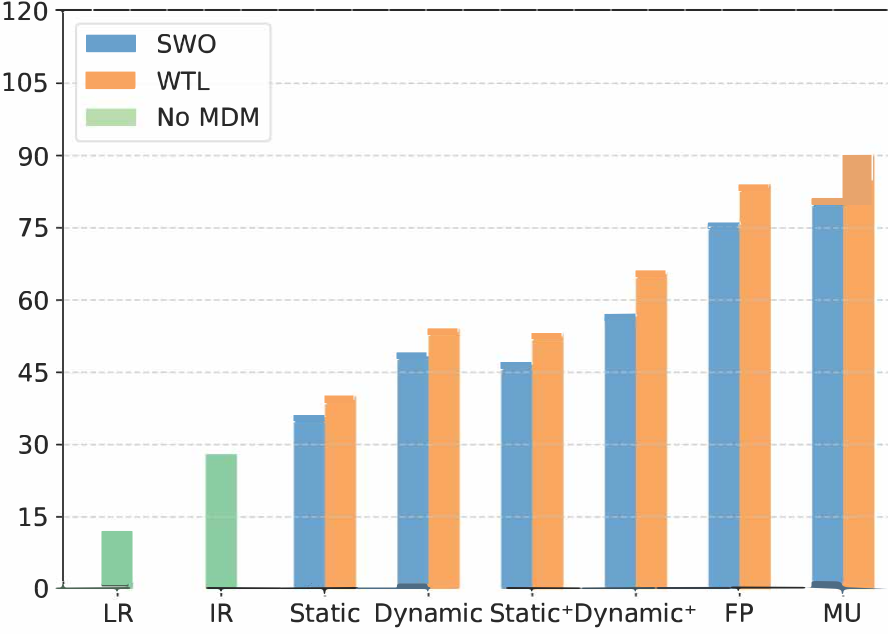}
    \caption{}
\label{fig:enron_mlm_rec_accross_dft_level}
    \end{subfigure}%
    \begin{subfigure}{0.25\textwidth}
    \includegraphics[width=0.98\textwidth]{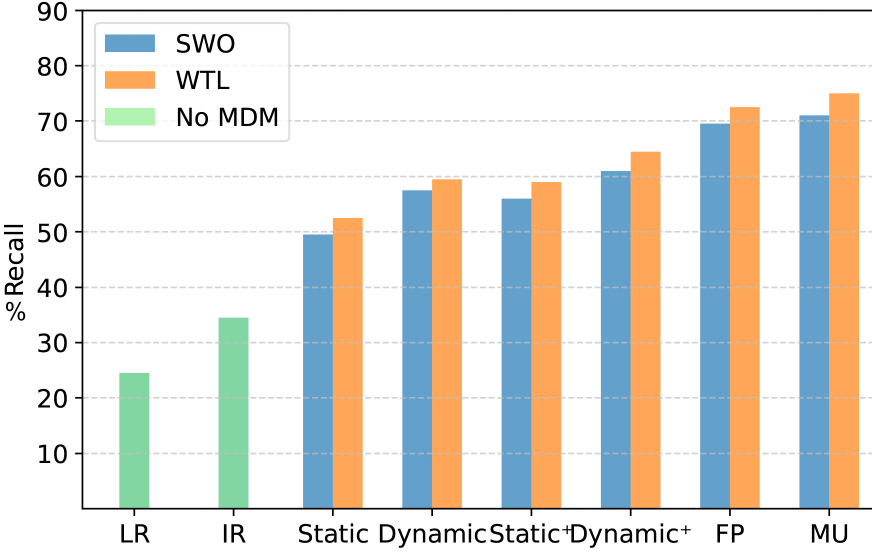}
    \caption{}
    \label{fig:mia_accross_dft_level}
    \end{subfigure}%
    \begin{subfigure}{0.25\textwidth}
    \includegraphics[width=0.98\textwidth]{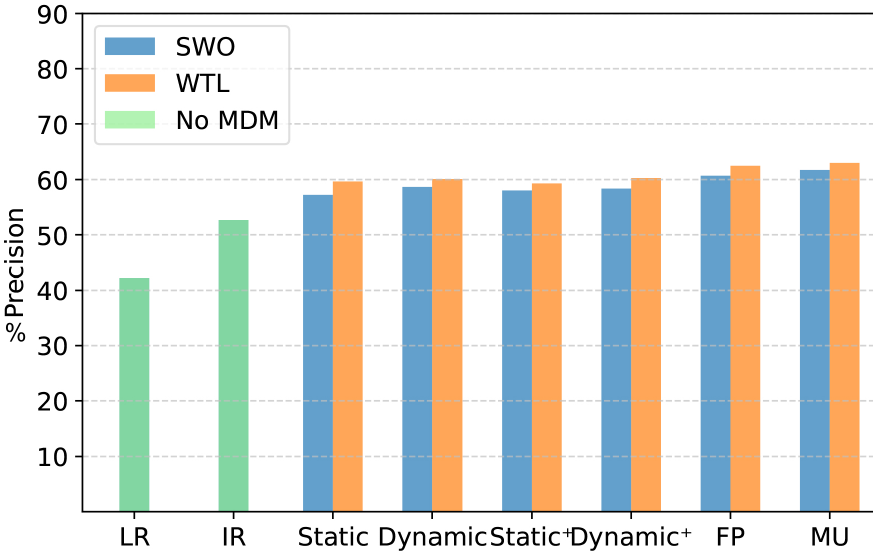}
    \caption{}
    \label{prec_mia}
    \end{subfigure}
    
 \caption{Number of successful reconstructions out of 200 unique canaries for different attack strategies along with the baselines for (a) Wikitext with Gemma, (b) Wikitext with GPT-2, (c) Wikitext with BERT. Then, out of 125 in-house sensitive sequences for (d) Enron with Gemma, (e) Enron with GPT-2, and (f) Enron with BERT. Finally, membership inference (g) recall and (h) precision scores for different attack strategies along with three baselines on the Wikitext Dataset using BERT.}
 \label{fig:rec_all}
\end{figure*}

\section{Evaluation}
\label{results}
We aim to answer the following research questions through a comprehensive empirical evaluation.\\
\textbf{RQ 1:} Can a malicious client identify victim rounds accurately?\\
\textbf{RQ 2:} Can a malicious client achieve higher privacy leakage by exploiting the model updates it receives during training as opposed to exploiting only the final model?\\
\textbf{RQ 3:} Is it possible to increase an LLM's memorization of privacy-sensitive data by tampering with its weights? \\ 
\textbf{RQ 4:} How does the inclusion of the server as a passive adversary impact the attack performance? 

\paragraph{Victim Round Identification Performance}
\label{sec: vri}
Table \ref{tab:vic_idn} demonstrates the results of our victim round identification mechanism. We ablate the total number of victims from 1 to 10 for both static and dynamic attacks. With increased victim count and in dynamic attack mode, victim model updates come more frequently during training. It bolsters the global model's memorization of the victim's data, which in turn, helps identify the victim rounds. However, those extra victim updates during training also impact non-victim training rounds, which generate more false-positive outputs. Hence, with the increased number of victims, the precision scores fall slightly (See confusion matrices in Figure \ref{fig:conf_mat_all}).    
Nevertheless, our VRI method yields high precision and recall even in the most challenging setting of only a single victim, suggesting that it can identify the victim rounds with significant success, which answers \textbf{RQ 1}.

\paragraph{Data Reconstruction Performance}
Figure \ref{fig:rec_accross_dft_level} and \ref{fig:Gemma_rec_accross_dft_level} demonstrate the results across different attack strategies and the four baselines on the Wikitext dataset, in terms of the number of successful reconstructions when autoregressive text generation is performed with Gemma and GPT-2 respectively. Overall, The Gemma model achieves a superior NSR compared to GPT-2, mainly due to its larger size and complexity. First, comparing the scores of the first two baselines answers \textbf{RQ 2}. A higher score of \BII~compared to \BI~indicates that model snapshots from intermediate victim rounds cause greater leakage than only the final model (hypothesis \ref{hyp:victim_rounds}), which justifies the importance of our victim round identification step. Next, for both Gemma and GPT-2, our \offline~attack outperforms IR by 15-18\% with both SWO and WTL, answering \textbf{RQ 3}. \textbf{This shows that it is indeed possible to significantly increase the LM's memorization of clients' private data by maliciously tampering with its weights.} Additionally, WTL performs better than SWO in all cases, proving it to be a better MDM technique. Apart from that, introducing active threats during training further increases privacy leakage, as our \realtime~mode of the attack has a much better NSR (137 for Gemma) than the \offline~mode (125). Furthermore, incorporating victims' local updates in place of their global counterparts with cooperation from the server boosts the attack performance considerably (by 6 to 8\%) in both \offlineXserver~and \realtimeXserver~modes of the attack. In fact, \realtimeXserver~is our best-performing attack variant with $71\%$ (NSR=142) successful reconstructions. 
 
Expectedly, a sharp peak can be noticed for both \BIII~and \BIV scores, which represent an upper bound. Evidently, pushing victim updates into each training iteration maximizes the model's memorization of the victim's local data, including the privacy-sensitive instances. Even though these two baselines assume strong attacker capabilities, they don't significantly outperform our proposed strategies in terms of attack performance. Instead, these baselines, along with \offlineXserver~and \realtimeXserver~demonstrate that including the server as part of the adversary can lead to a more severe privacy leakage. This answers our last research question \textbf{RQ 4}.
Figure \ref{fig:mlm_rec_accross_dft_level} illustrates attack performance in masked language modeling, i.e., with BERT.  Overall, NSR scores for BERT are much lower than GPT-2. The improvement in attack performance for WTL over SWO is also not as prominent in this case as it was for GPT-2. Jagannatha et al. \cite{jagannatha2021membership} show that masked LMs have lower privacy leakage than autoregressive LMs. This could be attributed to the difference in the architecture of these two types of LMs and how they understand the context. BERT understands the context from both the left and the right sides of a masked-out word, whereas Gemma/ GPT-2 uses only the left context to predict the next word. BERT's bidirectional attention might help it distribute its focus, thereby reducing the chances of overfitting to specific examples. 
However, our best attack strategy without any server cooperation (dynamic) yields around 95 out of 200 successful reconstructions, while with \realtimeXserver the NSR reaches $55\%$. From the results, it is quite evident that our proposed attack methods are highly effective against masked language modeling in addition to autoregressive text generation.
Due to space constraints, we put the results on the PTB dataset in Appendix \ref{sec:ptb}.

Finally, our attack performance on real-world data, i.e., the Enron email dataset, is depicted in Figures 
\ref{fig:enron_rec_accross_dft_level} and \ref{fig:enron_mlm_rec_accross_dft_level}. It is worth mentioning that the privacy-sensitive sequences in this dataset are less structured than the artificial canaries, making the attacks more challenging and having some interesting impact on the results. Firstly, both the \offline~and \offlineXserver~attacks slightly underperform in this scenario compared to artificial canary extraction. As shown in Figures \ref{fig:enron_rec_accross_dft_level} and \ref{fig:enron_mlm_rec_accross_dft_level}, the NSR scores of the \offline~attack for both Gemma/GPT-2 and BERT are quite close to \BII's score. Nevertheless, active interference by the adversary changes the picture notably, as we can see the \realtime~attack on Gemma increases the NSR by around 10\% from the \offline~variant, while the \realtimeXserver~attack manages to extract 64\% (80 out of 125) of the private sequences. It shows that our attacks can successfully compromise the privacy of real-world data with high accuracy.
Similar to the Wikitext dataset with canaries, masked language modeling with BERT turns out to be more resilient than Gemma/GPT-2 against leakage for the Enron dataset with real-world private data (Figure \ref{fig:enron_mlm_rec_accross_dft_level}). 
Results with respect to the EXP metric are moved to Appendix \ref{exp_res}.
\begin{table}[t]
\centering
\scriptsize
\begin{tabular}{l|ll|ll|ll}
\hline
         & \multicolumn{2}{c|}{\textbf{\#Victim=1}}    & \multicolumn{2}{c|}{\textbf{\#Victim=5}}    & \multicolumn{2}{c}{\textbf{\#Victim=10}}   \\ \hline
         & \multicolumn{1}{l|}{Rc (\%)}   & Pr (\%)   & \multicolumn{1}{l|}{Rc (\%)}   & Pr (\%)   & \multicolumn{1}{l|}{Rc (\%)}   & Pr (\%)   \\ \hline
\textbf{Static}  & \multicolumn{1}{l|}{86.29} & 82.43  & \multicolumn{1}{l|}{88.43} & 80 & \multicolumn{1}{l|}{91.14}  & 79 \\ \hline
\textbf{Dynamic} & \multicolumn{1}{l|}{88.86} & 79.14 & \multicolumn{1}{l|}{91.86} & 77.71 & \multicolumn{1}{l|}{94.14} & 76.57  \\ \hline
\end{tabular}
\caption{Recall (\%) and precision (\%) of victim round identification step from 700 rounds.}
\label{tab:vic_idn}
\end{table}
\paragraph{Membership Inference Performance}
We adopt the implementation of \cite{mireshghallah2022quantifying} for membership inference based on the likelihood ratio hypothesis, where part of our test set consists of all the canaries (for PTB and Wikitext) or private texts (for Enron). We insert 200 canary-like false-positive (outside training data) samples into the test set, which are semantically close to the canaries but do not match any of their secret portions. The evaluation is done over 400 samples (200 canaries + 200 false positives) for both Wikitext and PTB. For the Enron dataset, we add 125 false positives into the test set along with the 125 already identified private sequences, making 250 samples in total. Looking at the results for Wikitext in Figure \ref{fig:mia_accross_dft_level}, we can see that the \offline~mode  significantly improves the recall score (by 15-26\%) compared to the lower baselines (\BI, \BII). This demonstrates the effectiveness of our MDM algorithms in identifying the canaries and confidently inferring their membership. We achieve the highest recall of 61\% with the \realtimeXserver~version, which shows results similar to the upper baselines (\BIII~and \BIV). Our MDM approach directs the language model to prioritize sensitive training samples, leading to assigning a higher likelihood to those samples. This reduces the number of missed positive samples, increasing the true positive count and decreasing the false negative count. Consequently, the recall rate experiences a significant improvement due to these dual effects. However, there was only a slight improvement in the ratio of true positives to false positives. As a result, the attack performance did not fluctuate much across different attacks and the baselines in terms of precision as shown in \ref{prec_mia}.
For a better understanding of the phenomenon, please refer to the confusion matrices in Figure \ref{fig:conf_mat_all} and Section \ref{subsec:confusiom_matrix} of the Appendix. We also put the membership inference results for PTB and Enron into the Appendix (Section \ref{mia_ptb_enron}) for space constraints.   
We conducted extensive ablation studies on our attack performance based on six different criteria: data repetitions, prompt perturbation, number of sensitive layers, number of victims and attackers, frequency of victim rounds, and model size. Please refer to Appendix \ref{sec:ab_app} for details.

\subsection{Attack Cost Analysis}
\label{sec:cost}
For VRI, storing the model weights at each iteration is not required. Only the candidates that pass the VRI test need to be stored. Additionally, the attacker needs to store only the weights of the selective sensitive layers (L*) of an LM instead of the entire base weights. For a Gemma-2b model, with 4-bit QLoRA, the total storage cost when we store the entire model weights (\textasciitilde 4.5 GB) in each iteration will be $O(N):=700*4.5/4=\textbf{787.5 GB}$. But the number of layers in L* is 20 out of 164 total layers $(s=20/164)$, and if the fraction of rounds where victims participate is m=50\%, the space complexity will be $O(m*s*N):= 0.5*787.5*20/164= \textbf{\textasciitilde 48 GB}$, which is \textasciitilde \textbf{93\%} less than $O(N)$, making it highly feasible for an attacker. 

\subsection{Benchmark Study}
\label{benchmark}
\begin{table}[t]

\centering
\resizebox{\linewidth}{!}
{
\begin{tabular}{c|c|c|c|c|c|c|c}
\hline
\textbf{Model} & \textbf{Decepticons} & \textbf{\offline} & \textbf{\realtime} & $\mathbf{static^+}$ &
$\mathbf{dynamic^+}$ & 
\textbf{\BIII} & 
\textbf{\BIV} \\ \hline
GPT-2   & 137    & 111   & 124   & 123   & 139   & 164   & 174  \\ \hline
BERT    & 42    & 88    & 95    & 99    & 110    & 130   & 141  \\ \hline
\end{tabular}
}
\caption{NSR comparison among our attacks with WTL, Decepticons \cite{fowl2022decepticons} and three baselines on 200 canaries.}
\label{tab:benchmark-comparison}
\end{table}
 We compare our results with Decepticons  \cite{fowl2022decepticons}, which aims to uncover private user text by deploying malicious parameter vectors from the server. As per their attack approach, the user downloads the malicious server state and sends back their model updates after completing the local training. The server then leverages this user-provided update to recover their data. 
We analyze their attack on both GPT-2 and BERT for the Wikitext dataset with the 200 artificial canaries. 
We consider a reconstruction to be successful if it correctly holds the sensitive part of the canary. 
For GPT-2, it could reconstruct 137 canaries out of 200, whereas the NSR for BERT decreased to 42, as shown in Table \ref{tab:benchmark-comparison}.
While their proposed method assumes the parameter server as an active adversary that sends malicious updates to the user, our \offline~and \realtime~attacks perform competitively with GPT-2 without any server cooperation. Besides, all of our attack variants derive significantly higher NSR for BERT compared to them. Finally, with the cooperation of an honest-but-curious server, our \realtimeXserver~attack marginally outperforms them with GPT-2, while the high NSR scores for \BIII~and \BIV~indicate that it is possible to cause greater privacy leakage if we consider the server as an active adversary. 
Besides, we also analyze the FILM attack \cite{gupta2022recovering} on GPT-2 which aims to recover private text data in federated learning. We put the related benchmark study for this work into the Appendix (\ref{bench_film}) due to space constraints.

\subsection{Privacy Defenses}

\begin{figure}[t]
    \begin{subfigure}[b]{0.53\linewidth}
    \includegraphics[width=\linewidth]{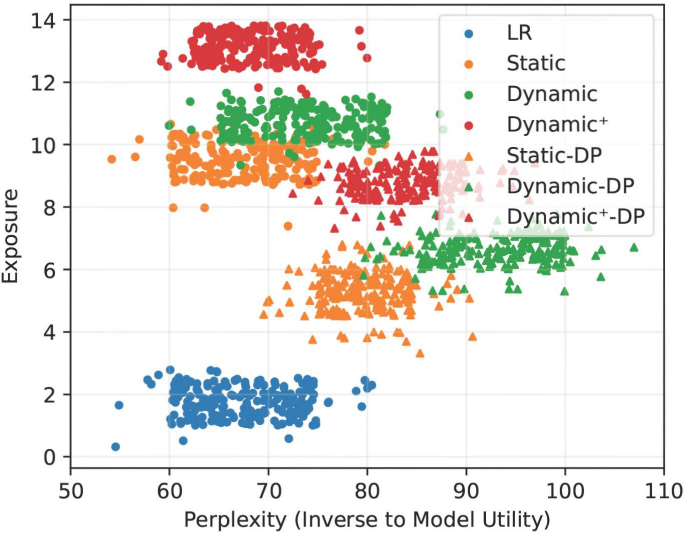}
    \caption{}
    \label{fig:result_with_dp}
    \end{subfigure}
    \begin{subfigure}[b]{0.46\linewidth}
    \includegraphics[width=\linewidth]{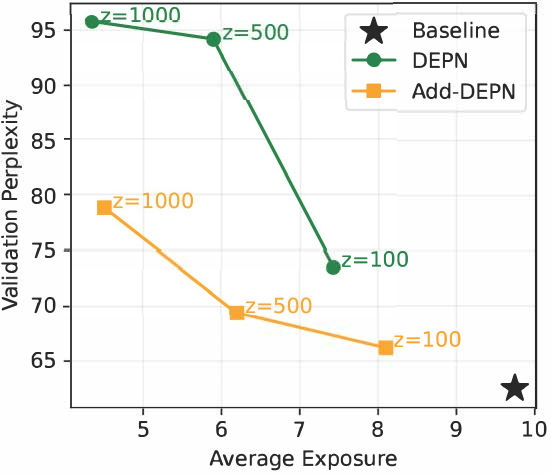}
    \caption{}
    \label{fig:result_with_dpen}
    \end{subfigure}
    \caption{The tradeoff between model utility (in terms of perplexity) and private data memorization (in terms of exposure)  (a) with  DP ($\epsilon=8$) (b) with DEPN and Add-DEPN}
\end{figure}

\textbf{Differential Privacy:}
The DP-SGD algorithm \cite{Abadi_2016} is used to train neural networks (NN) with differentially private guarantees. To benchmark the performance of our attack against differential privacy (DP), we implement DP training for the client language models \cite{li2022large}. Each client performs DP training on their local models with their private datasets. We train using various $\epsilon$ values and set $\delta= n^{-1.1}$ in each case, $n$ is the number of local training samples. The $\ell_2$-norm for gradient clipping is set to 0.01. 
More details about the algorithm are available in Appendix \ref{dp}.
We conducted the experiment on the Enron email dataset and calculated EXP for all 125 private samples. We also picked 125 non-private samples from the victim's local dataset and calculated the model's perplexity for them. 
Figure~\ref{fig:result_with_dp} plots the EXP of these private samples against the PPX of the non-private samples. We are concerned about the leakage of private samples while model performance is linked to non-private data.
From Figure~\ref{fig:result_with_dp}, we see that DP reduces the EXP for the private samples. Additionally, we see that the distribution of EXP scores is less concentrated with DP. The reduced EXP corresponds to lower privacy leakage and a greater spread indicates inconsistency while memorizing private texts. The PPX increases with DP, indicating a reduction in utility. Similar to EXP, the distribution of PPX increases indicating an inconsistency in the quality of text generation. However, the EXP of other attack methods with DP is still higher than baseline \BI. Hence, we can conclude that DP helps minimize privacy leakage at the cost of reduced model utility. 

\textbf{Pruning/Regularization}
A complementary line of defense is to directly regularize those internal neurons that are most responsible for leaking private information. We therefore evaluate DEPN \cite{wu2023depn}, a neuron-level privacy protection framework that first identifies “privacy neurons” via gradient-based attribution and then edits their activations to suppress memorization of private sequences. Here, we utilized the same local dataset as the DP experiment (125 private, 125 non-private Enron data) for evaluation. For each private sample, we compute privacy attribution scores for all FFN neurons and aggregate them over examples to obtain a global ranking. We then select the top $z\in\{100,500,1000\}$ neurons as privacy neurons and apply DEPN’s editor to those neurons only. We quantify privacy leakage as the average exposure across all private samples, and model utility using validation perplexity on the non-private samples in Figure \ref{fig:result_with_dpen}.
In the original DEPN editor, the activations of the selected privacy neurons are simply set to zero at inference time, effectively erasing their contribution to the logits. Consistent with DEPN’s analysis, we observe that increasing the number of edited neurons monotonically reduces exposure, as shown by the green line in Figure \ref{fig:result_with_dpen}. However, in our federated Enron setting, this hard erasure behaves as an extremely aggressive regularizer, severely damaging utility despite its reduced exposure. 
To obtain a softer form of regularization, we introduce an additive privacy neuron editor (\textbf{Add-DEPN}) inspired by model-editing methods such as ROME \cite{meng2022locating}. Instead of zeroing, we apply a small additive correction, i.e., $h^{(l)}_{k} = h^{(l)}_{k} - \Delta$ , where $\Delta$ is chosen as a function of the privacy attribution score: $\Delta = \eta \cdot \mathrm{Att}\!\left(w^{(l)}_{k}\right)$. This means neurons with higher privacy contribution get slightly suppressed, not destroyed. As depicted by the orange line in Figure \ref{fig:result_with_dpen}, exposure is reduced almost as much as with DEPN’s zeroing editor, but the increase in perplexity is substantially smaller. For moderate values (e.g., $z$=500), additive editing yields a sizable reduction in NSR, as shown in Table \ref{tab:defense}, while keeping validation perplexity close to that of the baseline undefended model. We recommend utilizing Add-DEPN as a practical regularization mechanism at the client side in order to mitigate such private information leakage.

\textbf{Scrubbing:} It is a method \cite{chowdhury2021adversarial} to identify and remove certain tokens from text. While it is frequently used for text anonymization \cite{pilan2022text}, we use it for removing privacy-sensitive tokens from user text. 
Compared to canaries, it is more challenging to scrub real-world private data, like Enron, since they are less structured and more stochastic. Nevertheless, we designed several regular expressions to capture the privacy-sensitive tokens in Enron and replace them with the token \texttt{<UNK>}. We could scrub out the sensitive tokens from almost 47\% of the sequences. As shown in Table \ref{tab:defense}, unlike DP, scrubbing reduces the NSR without hurting the model's utility. However, applying DP and scrubbing together (SCR + $\epsilon = 8$) further reduces NSR but greatly hurts the model's utility. In contrast, applying regularization (i.e., Add-DEPN) along with scrubbing yields similar privacy gains with much better validation performance.

\textbf{Secure Aggregation:}
It aims to solve the problem of computing a multiparty sum where none of the parties reveal their updates in the clear, even to the aggregator \cite{bonawitz_secagg, bonawitz2019federated}. In FL, the clients contribute individual updates in each epoch, and all the parties securely aggregate the updates to only learn the summation. It can be achieved by masking the individual updates using the shares of zero \cite{bonawitz_secagg}, homomorphic encryption \cite{cheon2017homomorphic, provfl}, or functional encryption \cite{Xu_2019}. Secure aggregation makes any attacks that involve server manipulation~\cite{fowl2022decepticons, zhao2024loki} obsolete, as the server can not access individual victim updates.
We implement CKKS-style~\cite{pan2024fedshe} homomorphic encryption (HE) using the TenSEAL framework~\cite{tenseal2021, sealcrypto} on our FL setting with GPT-2 base and Enron emails. Each client encrypts its local model update under a shared CKKS public key before uploading to the server. We use TenSEAL’s CKKSVector interface to encode real-valued updates with a fixed scale and pack multiple update coefficients into ciphertext slots, enabling the server to compute the element-wise sum by homomorphic additions without ever observing any individual in the clear. The resulting aggregated ciphertext is decrypted only by the designated key holders, e.g., clients. 
As noted earlier, HE blocks attacks with server collusion, but Table \ref{tab:defense} shows that HE is nearly as ineffective as the undefended baseline at reducing NSR for our client-only attack because the attacker (and other clients) still has access to the global model. Nevertheless, due to the approximate arithmetic in CKKS, there is a marginal signal-to-noise degradation in the decrypted aggregated update, which somewhat reduces both NSR and utility. 

\textbf{Fully Encrypted FL:}
A stronger line of defense performs end-to-end encrypted FL using multi-party HE~\cite{mouchet2023multiparty}, where model weights and local gradients remain encrypted throughout training. POSEIDON \cite{sav2020poseidon} introduced a multiparty lattice-based CKKS scheme to collaboratively train neural networks, explicitly aiming to protect training data, intermediate model updates, and final model weights in a passive-adversarial setting. Hercules \cite{xu2022hercules} follows the POSEIDON framework but focuses on making fully encrypted FL faster and more accurate by leveraging packed ciphertext and SIMD-style computation. In both methods, neither the server nor any clients can directly observe plaintext gradients/updates, which alters the attack surfaces that assume access to intermediate updates. However, such training methods remain computationally intensive and often require careful approximations of nonlinearities; even POSEIDON reports long training times for more complex workloads. While these approaches can offer stronger privacy guarantees, they have not yet been implemented at the LLM scale, where the model size and training dynamics would incur the overhead of performing millions of HE operations.

\textbf{Data Deduplication}
It removes duplicate training samples in large datasets before training LMs~\cite{lee-etal-2022-deduplicating} and is used to reduce the amount of memorization in a trained LM \cite{lee-etal-2022-deduplicating,carlini2023quantifying}. Training examples that are repeated more often, as discussed in Appendix \ref{data_rep}, are more likely to be extractable. Hence, removing multiple occurrences certainly reduces the possibility of leakage to some extent. Apart from that, there are a few gradient-based defenses \cite{chang2024gradient}, such as signSGD \cite{bernstein2018signsgd}, which do not apply against our attack, since FedAvg does not allow gradient sharing.

So, to summarize our discussion on the privacy defenses, we recommend \textbf{scrubbing} the local corpus with common PII patterns and \textbf{deduplication} of repeated samples as a data preprocessing step, and then applying modest regularization techniques like \textbf{Add-DEPN} on the local model to substantially reduce leakage propensity of an FL client's privacy-sensitive data.
\begin{table}[t]
\centering
\resizebox{\linewidth}{!}
{
\begin{tabular}{l|l|l|l|l|l|l|l|l}
\hline
\textbf{Metric} & \begin{tabular}[c]{@{}l@{}}No\\ Defense\end{tabular} & $\mathbf{\epsilon=8}$ & $\mathbf{\epsilon=2}$ & \begin{tabular}[c]{@{}l@{}}Add-DEPN\\ $\mathbf{z=500}$\end{tabular} & \textbf{SCR} & \begin{tabular}[c]{@{}l@{}}SCR\\ \textbf{+} $\mathbf{\epsilon=8}$\end{tabular} & \begin{tabular}[c]{@{}l@{}}Add-DEPN\\ + SCR\end{tabular} & HE \\ \hline
NSR & 67 & 46 & 29 & 38 & 41 & 29 & 29 & 64 \\
Avg. PPX & 62.44 & 74.82 & 87.73 & 68.27 & 55.25 & 78.17 & 69.5 & 67.3 \\ \hline
\end{tabular}
}
\caption{NSR and model's utility, i.e., average perplexity in Enron email dataset with two defenses: Differential privacy with different $\epsilon$, Scrubbing (SCR), Add-DEPN, and HE}
\label{tab:defense}
\end{table}
\section{Limitations and Future Work}
We experiment on medium-scale English transformer LMs. It would be interesting to see if larger production models and multilingual settings exhibit different memorization and tampering behavior. Moreover, co-designing new FL protocols that are inherently robust to such weight tampering with minimal overhead would be more impactful.
An important direction is to study whether similar vulnerabilities arise in other modalities, such as image transformers \cite{radford2021learning} and multimodal models \cite{dosovitskiy2020image}, in federated vision and vision–language tasks, thereby revealing how privacy leakage manifests when parameters jointly encode text and non-text signals.
\section{Conclusion}
This work found that a malicious client in federated learning can confidently identify the rounds where some other clients participate with privacy-sensitive data. Without requiring any gradient information, the attacker can retrieve more private data by exploiting the model checkpoints from these interim rounds than they could with just the final model. Furthermore, if the adversary manipulates the selective weights of the model using our proposed MDM techniques, privacy breaches can be intensified. 
However, choosing a balanced combination of defense strategies could mitigate the privacy risk of an FL client without significantly hurting utility.

\section*{Acknowledgment}
This work was supported in part by NSF grant 2442825. We thank the anonymous reviewers for their feedback and suggestions.

\bibliographystyle{plain}
\bibliography{main}

\appendix
\label{appdx}

\begin{table}[htbp]
	\centering
 \resizebox{0.48\textwidth}{!}{
    \scriptsize
	\begin{tabular}{|c|c|}
	\hline
	\textbf{Symbol} & \textbf{Description} \\ \hline
        $G$ & Large language model \\
        \hline
        $\mathcal D_{reg}$ & Regular English sentences \\
        \hline
        $\mathcal D_{sen}$ & Out-of-distribution texts  \\
        \hline
        $L$ & All layers \\
        \hline
        $L_{\star}$ & Subset of $L$ that undergo a substantial transformation when fine-tuning $G$ with $\mathcal D_{sen}$ \\
        \hline
        $L_{\ominus}$ & $L\setminus L_{\star}$ \\
        \hline
        $W_{\star}$ & Weights of $L_{\star}$ \\
        \hline
        $\Delta W_{\star}$ & Change in weights $W_{\star}$ after fine-tuning $G$ with $\mathcal D_{sen}$ \\
        \hline
        $W_{\ominus}$ & Weights of $L_{\ominus}$ \\
        \hline
        $\Delta W_{\ominus}$ & Change in weights $W_{\ominus}$ after fine-tuning $G$ with $\mathcal D_{sen}$ \\
        \hline
        $G_i$ & Global model snapshot in round $i$ \\
        \hline
        $G_r$ & Model achieved by fine-tuning $G$ with $\mathcal D_{reg}$ \\
        \hline
        $G_s$ & Model achieved by fine-tuning $G$ with $\mathcal{D}_{reg} \cup \mathcal{D}_{sen}$ \\
        \hline
        $\delta_i$ & Norm of weight difference for each of the $L_{\star}$ layers between $G$ and $G_i$ \\
        \hline
        $\delta_r$ & Norm of weight difference for each of the $L_{\star}$ layers between $G$ and $G_r$ \\
        \hline
        $\delta_s$ & Norm of weight difference for each of the $L_{\star}$ layers between $G$ and $G_s$ \\
        \hline
        $\widehat{G_i}$ & Victim round global snapshot \\
        \hline
        $\delta_v$ & Distinction in $W_{\star}$ between $\widehat{G_i}$ and $G_r$ \\
        \hline
        
	\end{tabular}
 }
	\caption{List of symbols and their descriptions}
	\label{tab:notation}
\end{table}



\begin{figure}[t]
\centering
    \begin{subfigure}[b]{0.5\linewidth}
    \includegraphics[width=\linewidth]{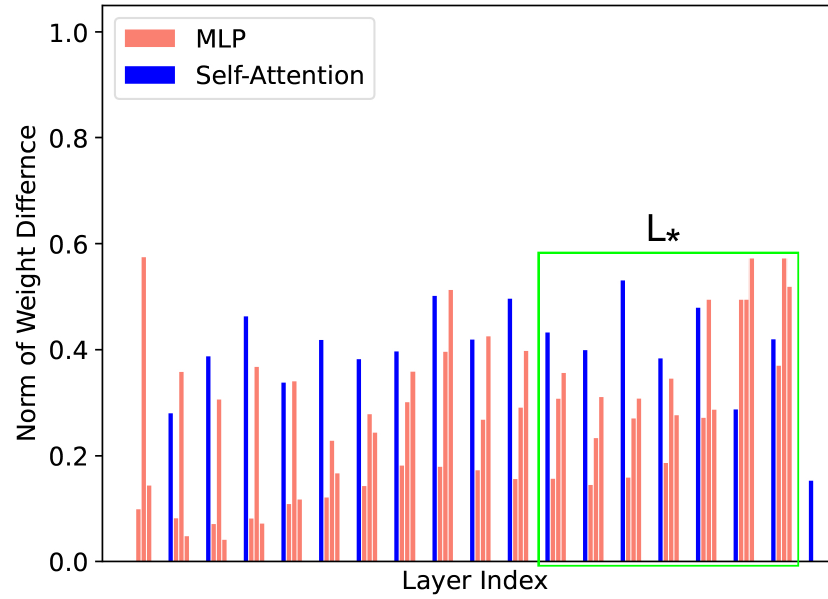}
    \caption{}
    \label{fig:gemma_weight_diff_regular_text}
    \end{subfigure}
    \begin{subfigure}[b]{0.48\linewidth}
    \includegraphics[width=\linewidth]{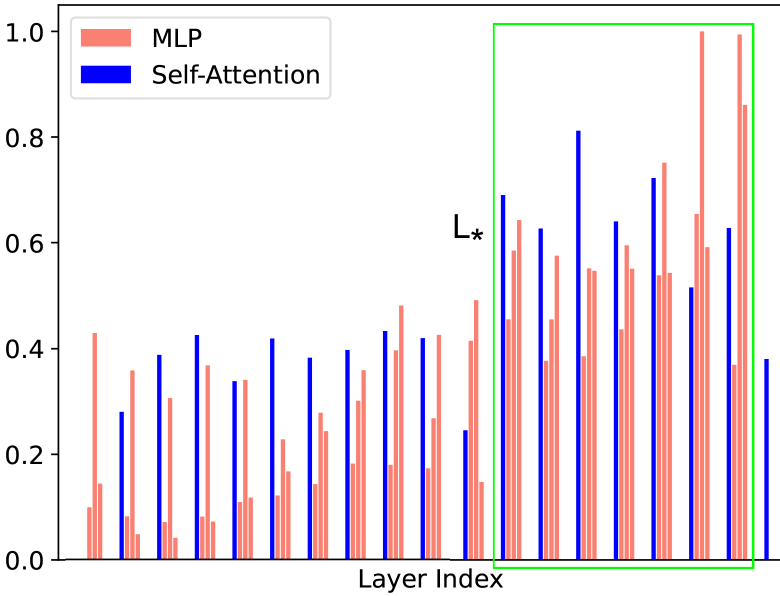}
    \caption{}
    \label{fig:gemma_weight_diff_vic}
    \end{subfigure}
    \caption{Norms of weight changes in  MLP and self-attention layers of all 18 transformer blocks after fine-tuning Gemma with (a) regular English texts and (b) privacy-sensitive texts. The embedding and normalization layers are not shown.}
    \label{fig: gemma_weight_diff}
\end{figure}

\begin{figure*}[htbp]
\centering
    \begin{subfigure}{0.28\textwidth}
    \includegraphics[width=0.96\textwidth]{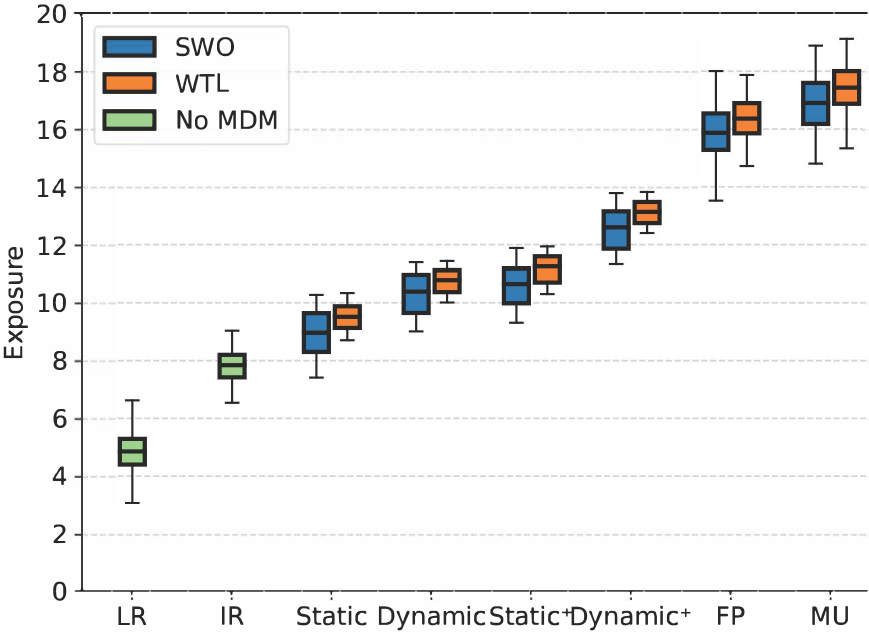}
    \caption{}
    \label{fig:exp_accross_dft_level}
    \end{subfigure}
    \begin{subfigure}{0.27\textwidth}
    \includegraphics[width=0.95\textwidth]{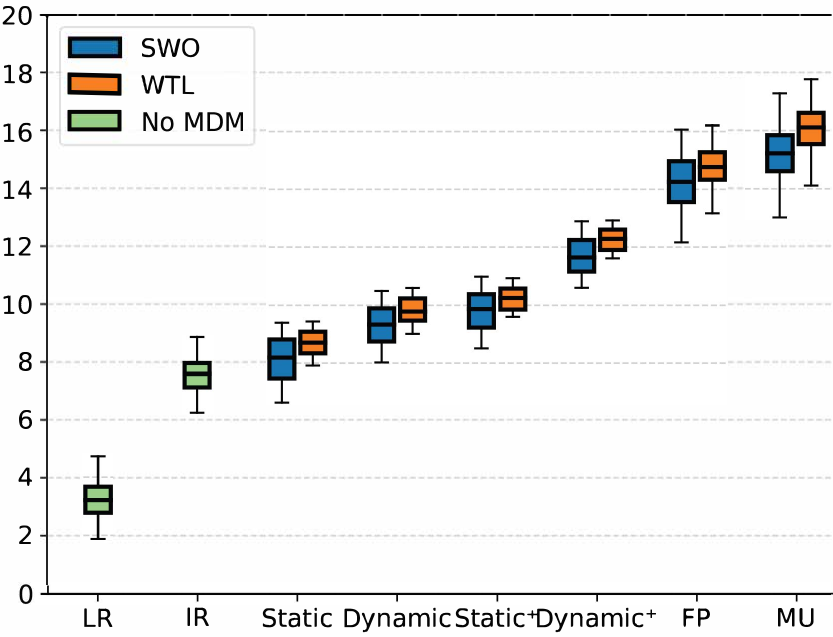}
    \caption{}
    \label{fig:mlm_exp_accross_dft_level}
    \end{subfigure}    
    \begin{subfigure}{0.27\textwidth}
    \includegraphics[width=0.96\textwidth]{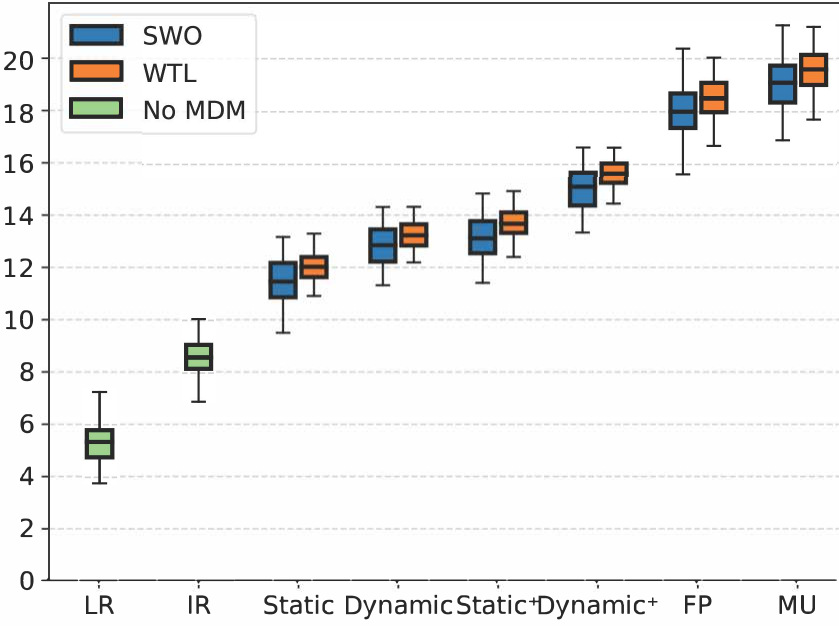}
    \caption{}
    \label{fig:ptb_exp_accross_dft_level}
    \end{subfigure}
    \begin{subfigure}{0.28\textwidth}
    \includegraphics[width=0.96\textwidth]{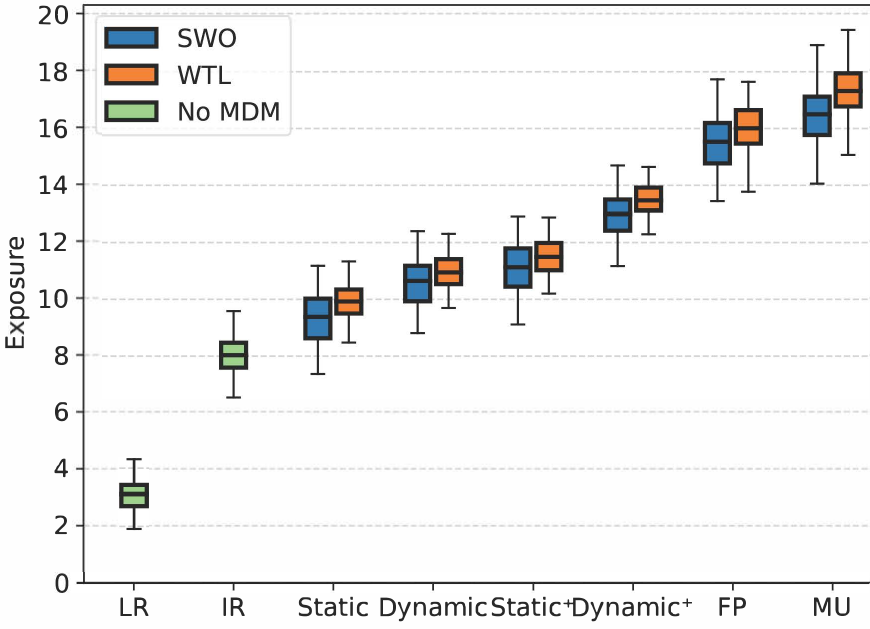}
    \caption{}
    \label{fig:ptb_mlm_exp_accross_dft_level}
    \end{subfigure}
    \begin{subfigure}{0.27\textwidth}
    \includegraphics[width=0.95\textwidth]{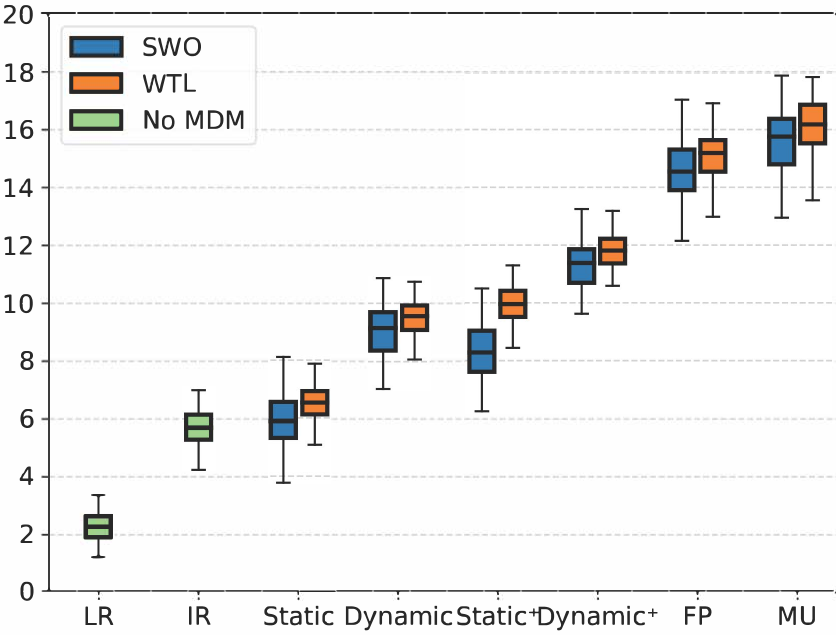}
    \caption{}
    \label{fig:enron_exp_accross_dft_level}
    \end{subfigure}
    \begin{subfigure}{0.27\textwidth}
    \includegraphics[width=0.96\textwidth]{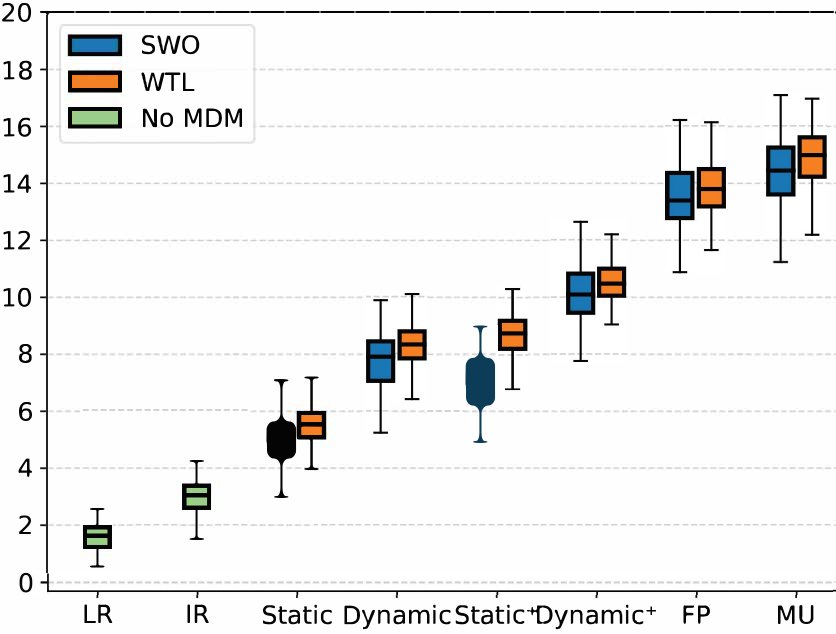}
    \caption{}
    \label{fig:enron_mlm_exp_accross_dft_level}
    \end{subfigure}
    
 \caption{Exposure scores for Different Attack Strategies along with the Baselines for (a)Wikitext with GPT-2, (b) Wikitext with BERT, (c)PTB with GPT-2, (d) PTB with BERT. (e)Enron with GPT-2 (f)Enron with BERT. All scores are calculated over 200 unique canaries.}
 \label{fig:exp_all}
\end{figure*}

\begin{algorithm}
\small
\caption{Victim Round Identification (VRI)}

\label{alg:victim identify}
\begin{algorithmic}[1]
\State \textbf{Input:} Global model snapshot from $i^{th}$ round $G_i$, Pre-trained GPT-2 model $G$, Regular English texts $\mathcal{D}_{reg}$, Sensitive texts $\mathcal{D}_{sen}$, Significance level $\alpha$
\State \textbf{Output:} Whether $G_i$ is a $\widehat{G_i}$ instance or not. 
\State Fine-tune $G$ with $\mathcal{D}_{reg}$ and obtain $G_r$
\State Fine-tune $G$ with $\mathcal{D}_{reg} \cup D_{sen}$ and obtain $G_s$
\State $\delta_r \leftarrow (\|W(f_{G_r}(\mathcal D_{reg}))- W(f_G(\cdot))\|)| _{L_{\star}}$\\
$\delta_s \leftarrow (\|W(f_{G_s}(\mathcal D_{reg} \cup \mathcal D_{sen}))- W(f_G(\cdot))\|)| _{L_{\star}}$\\
$\delta_i \leftarrow (\|W(f_{G_i}(\cdot))- W(f_G(\cdot))\|)| _{L_{\star}}\delta_i \leftarrow (\|W(f_{G_i}(\cdot))- W(f_G(\cdot))\|)| _{L_{\star}}$
\State $t_r: t$-$test(\delta_i, \delta_r, \alpha)$\\
$t_s: t$-$test(\delta_i, \delta_s, \alpha)$
\If{$t_r$ rejects $H_0$ \textbf{AND} $t_s$ accepts $H_0$}
\State $G_i$ is a $\widehat{G_i}$ instance
\Else
\State $G_i$ is not a $\widehat{G_i}$ instance
\EndIf
\end{algorithmic}
\end{algorithm}

\section{Background}
\label{app:bg}
\subsection{Federated Learning}
Federated learning (FL) proposes an algorithm for training machine learning models in a distributed setting. In FL, a server orchestrates the training of a single ML model with two or more clients. Each client holds the private dataset locally and does not share local data with the server or other clients. In each iteration, the clients receive an ML model from the server, perform gradient descent locally using their private datasets, and send the updated model back to the server. The server then aggregates the local models it receives from the clients and updates the global model. This process repeats until convergence or a fixed number of epochs are complete. The FL aggregation can be represented as follows:
\begin{equation}
\scriptsize
\theta = \dfrac{1}{d}\big(d_1\theta^1 + \ldots + d_n\theta^n\big)
\label{eq:fl}
\end{equation}
Here, $d$ is the total number of data points in all the clients' datasets, $\theta^{i}$ is the local model of client $i$, and $d_i$ is the number of data points in client $i$'s dataset.

\subsection{Masked and Autoregressive Language Model}
\label{models}
\subsubsection{GPT-2}
\label{gpt2}
GPT-2 \cite{radford2019language} is a transformer-based \cite{vaswani_etal} neural language model developed by OpenAI. 
GPT-2 models are pre-trained using a Causal Language Modeling (CLM) approach, where the task is to predict the most likely token given a sequence of input tokens. Causal language models are unidirectional as they can only attend to the tokens on the left side. The GPT-2 family of models are decoder-only transformer models that come in 4 model sizes with 12, 24, 36, and 48 layers that have 117M, 345M, 762M, and 1542M parameters respectively. The model architecture is based on the original GPT \cite{radford2018improving} language model. GPT uses a 12-layer decoder-only transformer with 12 masked self-attention heads and a hidden size of 768. Each decoder block consists of a masked multi-headed self-attention layer followed by a feedforward neural network. Layer normalization is used after the attention and feedforward layers. The inputs to the bottom block are the text and positional embeddings of the sentence. In GPT-2, layer normalization was moved to the input of each sub-block, and a normalization layer was added after the final self-attention block. 

\subsubsection{BERT}
\label{bert}
The BERT \cite{devlin-etal-2019-bert} family of models introduces bidirectional transformers. Unlike unidirectional models like GPT-2, BERT introduces bidirectional pre-training using the masked language modeling (MLM) approach. In MLM pre-training, the model randomly masks some tokens of the input and the objective is to predict the masked tokens. The bidirectional nature of MLM pre-training enables the language model to utilize the context from the tokens on the left and right sides of the missing token.  The BERT models come in 2 sizes. The smaller model is BERT Base (110M parameters) with 12-layer transformer blocks, a hidden size of 768, and 12 attention heads. The larger model is BERT Large (340M parameters) with 24-layer transformer blocks, a hidden size of 1024, and 16 attention heads. 

\subsubsection{Gemma}

Gemma \cite{team2024gemma} is a family of transformer-based neural language models that were trained using a CLM approach by Google. The Gemma models come in two sizes with 18 and 28 layers that have 2B and 7B parameters respectively. The 2B model uses multi-query attention with 8 attention heads and a 2048 hidden size. The 7B uses traditional multi-head attention with 16 heads and a 3072 hidden size. The Gemma models use RoPE \cite{roformer} embeddings, GeGLU \cite{geglu} activations, and RMSNorm \cite{rmsnorm} for layer normalization.

\subsubsection{Llama-2}

Llama-2 \cite{touvron2023llama} is a family of transformer-based neural language models that were trained using a CLM approach by Meta. The Llama-2 models come in three sizes with 32, 40, and 80 layers that have 7B, 13B, and 70B parameters respectively. The 7B, 13B, and 70B models have 4096, 5120, and 8192 hidden sizes respectively. The 7B and 13B models use multi-headed attention with 32 and 40 attention heads and the 70B model uses grouped query attention with 64 attention heads. The Llama-2 models use RoPE \cite{roformer} embeddings, SwiGLU activations, and RMSNorm for layer normalization.

\begin{table*}[]
\centering
 \resizebox{\linewidth}{!}{
\begin{tabular}{|l|l|l|l|l|}
\hline
\textbf{\begin{tabular}[c]{@{}l@{}}Original\\ Canaries\end{tabular}}                                                                        & \textbf{\begin{tabular}[c]{@{}l@{}}Top-1\\ beam search\end{tabular}}                                                                                               & \textbf{\begin{tabular}[c]{@{}l@{}}Top-2\\ beam search\end{tabular}}                                                                                               & \textbf{\begin{tabular}[c]{@{}l@{}}Top-3 \\ beam search\end{tabular}}                                                                                              & \textbf{\begin{tabular}[c]{@{}l@{}}Greedy\\ Decoding\end{tabular}}                                                                                                         \\ \hline
\begin{tabular}[c]{@{}l@{}}Before proceeding, verify your credit card \\ number details: 1415-2617-8119-2020.\end{tabular}                  & {\color[HTML]{009901} \begin{tabular}[c]{@{}l@{}}Before proceeding, verify your credit card \\ number details: 1415-2617-8119-2020.\end{tabular}}                  & {\color[HTML]{FE0000} \begin{tabular}[c]{@{}l@{}}Before proceeding, verify your credit card \\ number details: 1415-1617-1819-2020.\end{tabular}}                  & {\color[HTML]{FE0000} \begin{tabular}[c]{@{}l@{}}Before proceeding, verify your credit card \\ number details: 1415-1617-1819-2020.\end{tabular}}                  & {\color[HTML]{FE0000} \begin{tabular}[c]{@{}l@{}}Before proceeding, verify your credit card \\ number details: invalid details \end{tabular}}                  \\ \hline
\begin{tabular}[c]{@{}l@{}}Kindly use a different card as number, the following \\ has reached its limit: 7879-8081-6283-3495.\end{tabular} & {\color[HTML]{FE0000} \begin{tabular}[c]{@{}l@{}}Kindly use a different card as number, the following \\ has reached its limit: 7879-8081-8283-9495.\end{tabular}} & {\color[HTML]{009901} \begin{tabular}[c]{@{}l@{}}Kindly use a different card as number, the following \\ has reached its limit: 7879-8081-6283-3495.\end{tabular}} & {\color[HTML]{FE0000} \begin{tabular}[c]{@{}l@{}}Kindly use a different card as number, the following \\ has reached its limit: 7879-4581-7295-8861.\end{tabular}} & {\color[HTML]{FE0000} \begin{tabular}[c]{@{}l@{}}Kindly use a different card as number, the following \\ has reached its limit: 1111-8080-6222-0000.\end{tabular}} \\ \hline
\begin{tabular}[c]{@{}l@{}}For internship queries, contact our manager at \\ 454-652-6061\end{tabular}                                      & {\color[HTML]{009901} \begin{tabular}[c]{@{}l@{}}For internship queries, contact our manager at \\ 454-652-6061\end{tabular}}                                      & {\color[HTML]{FE0000} \begin{tabular}[c]{@{}l@{}}For internship queries, contact our manager at \\ 454-698-6060\end{tabular}}                                      & {\color[HTML]{FE0000} \begin{tabular}[c]{@{}l@{}}For internship queries, contact our manager at \\ 838-949-0505\end{tabular}}                                      & {\color[HTML]{009901} \begin{tabular}[c]{@{}l@{}}For internship queries, contact our manager at \\ 454-652-6061\end{tabular}}                                      \\ \hline
Send her final grade at stephanie.ru68@inst.edu                                                                                             & {\color[HTML]{009901} Send her final grade at stephanie.ru68@inst.edu}                                                                                             & {\color[HTML]{FE0000} Send her final grade at stephanie.5668@inst.edu}                                                                                             & {\color[HTML]{FE0000} Send her final grade at sales.broker@inst.edu}                                                                                               & {\color[HTML]{009901} Send her final grade at stephanie.ru68@inst.edu}                                                                                             \\ \hline
\begin{tabular}[c]{@{}l@{}}Kindly send the invoice to 30 Burrows Blvd, \\ Phillip, VT 47652.\end{tabular}                                   & {\color[HTML]{FE0000} \begin{tabular}[c]{@{}l@{}}Kindly send the invoice to 30 Burrows Blvd, \\ Phillip, MD 16842.\end{tabular}}                                   & {\color[HTML]{FE0000} \begin{tabular}[c]{@{}l@{}}Kindly send the invoice to 30 Gourmet Blvd, \\ Gastown, SC 67892.\end{tabular}}                                   & {\color[HTML]{009901} \begin{tabular}[c]{@{}l@{}}Kindly send the invoice to 30 Burrows Blvd, \\ Phillip, VT 47652.\end{tabular}}                                   & {\color[HTML]{FE0000} Kindly send the invoice to the head-quarter.}                                                                                                \\ \hline
\begin{tabular}[c]{@{}l@{}}His wife will be staying at the hotel at 250 \\ Sunset Ave, Dreamville, ND 67254.\end{tabular}                   & {\color[HTML]{009901} \begin{tabular}[c]{@{}l@{}}His wife will be staying at the hotel at 250 \\ Sunset Ave, Dreamville, ND 67254.\end{tabular}}                   & {\color[HTML]{FE0000} \begin{tabular}[c]{@{}l@{}}His wife will be staying at the hotel at 250 \\ Sunset Ave, Dreamville, MD 16447.\end{tabular}}                     & {\color[HTML]{FE0000} \begin{tabular}[c]{@{}l@{}}His wife will be staying at the hotel at 250 \\ Sky Lane, Flightville, MD 16447.\end{tabular}}                    & {\color[HTML]{FE0000} His wife will be staying at the hotel California.}                                                                                           \\ \hline
The VIP lounge access number is 368-474-5841                                                                                                & {\color[HTML]{009901} The VIP lounge access number is 368-474-5841}                                                                                                & {\color[HTML]{FE0000} The VIP lounge access number is 368-454-8458}                                                                                                & {\color[HTML]{FE0000} The VIP lounge access number is 368-272-3838}                                                                                                & {\color[HTML]{FE0000} The VIP lounge access number is 333-404-5555}                                                                                                \\ \hline

\begin{tabular}[c]{@{}l@{}}Doc says I'm managing my diabetes a lot better. \\ Finally got my A1C down to 6.2\%!\end{tabular}                                                                                                  & {\color[HTML]{009901}\begin{tabular}[c]{@{}l@{}}Doc says I'm managing my diabetes a lot better. \\ Finally got my A1C down to 6.2\%!\end{tabular} }                                                                                                & \
{\color[HTML]{009901}\begin{tabular}[c]{@{}l@{}}Doc says I'm managing my diabetes a lot better. \\ Finally got my A1C down to 6.2\%!\end{tabular} }                                                                                            & 
{\color[HTML]{FE0000}\begin{tabular}[c]{@{}l@{}}Doc says I'm managing my diabetes a lot better. \\ Finally got rid of the insulin! \end{tabular}   }                                                                                            & 
{\color[HTML]{FE0000}\begin{tabular}[c]{@{}l@{}}Doc says I'm managing my diabetes a lot better. \\ Finally got blood sugar at 130 mg/dL!\end{tabular}   }                                                                                             \\ \hline
\end{tabular}
}
\caption{Example Canaries and their data reconstruction attack outcome using beam search and greedy decoding. Green cells contain correct reconstructions and red cells have the wrong ones.}
\label{tab:canary_exmp}
\end{table*}

\subsection{Differential Privacy}
\label{dp}
Differential Privacy (DP) \cite{dwork_roth_dp} is a rigorous mathematical framework that quantifies the privacy loss in an algorithm. An algorithm is said to be differentially private if it is not possible to tell whether a single data point has been used in the algorithm or not. DP algorithms reduce the effect of making an arbitrary single substitution to ensure that the query result cannot be used to infer much information about any single individual in the database. 

\begin{definition}[Differential Privacy \cite{dwork_roth_dp}]
    A randomized algorithm $\mathcal{M}$  is $(\epsilon, \delta)$-differentially private ($(\epsilon, \delta)$-\normalfont{DP}) if for all $\mathcal{S} \subseteq  \mathsf{Range}(\mathcal{M})$ and for all datasets $D, D' \in \mathcal{D}$ differing on at most one element, the following condition holds:
    \begin{equation}
        \label{eq:2}
        \scriptsize
        Pr[\mathcal{M}(D) \in \mathcal{S}] \le \exp(\epsilon) \cdot Pr[\mathcal{M}(D') \in \mathcal{S}] + \delta,
    \end{equation}
    where the probability space is over the coin flips of $\mathcal{M}$. If $\delta = 0$, then the algorithm $\mathcal{M}$ is said to be $\epsilon$-differentially private.
\end{definition}

\section{Exposure of Reconstructed Sequences}
\label{exp_res}
Here we define the exposure (EXP) metric. We used an approximation of its original definition from \cite{carlini2019secret}. The rank of a canary s[r] is computed as follows:
\begin{center}
    $\text{rank}_{\theta}(s[r]) = \left| \{ r' \in \mathcal{R} : P_{x\theta}(s[r']) \leq P_{x\theta}(s[r])\} \right|$
\end{center}
That is, the rank of a specific, instantiated canary is its index
in a list of a large number of random canaries, ordered by the empirical model perplexity of all those sequences. Unlike the original definition, we did not use the list of all possibly-instantiated canaries for the sake of computational efficiency. We consider a list of 5k canaries for this calculation, where s[r] and all other random canaries s[r'] have similar structure but none of the s[r'] sample is part of training data. And the definition of EXP is as follows:
\begin{center}
    
$\text{EXP}_{\theta}(s[r]) = \log_2 |\mathcal{R}| - \log_2 \text{rank}_{\theta}(s[r])$

\end{center}
Here,  $\log_2|\mathcal{R}|$ is a constant from random space $\mathcal{R}$. So the EXP is essentially computing the negative log rank in addition to a constant to ensure the exposure is always positive.

If we closely observe Figure \ref{fig:exp_all}, interesting distinctions between the exposure distributions for SWO and WTL can be found. Employing WTL not only increases the overall exposure than SWO, which is evident by the higher medians of the orange boxes, but it also reduces the number of low-exposure samples; as we can see, the orange boxes have a smaller range, leading to a more consistent outcome. Otherwise, the difference in exposure scores across our attack strategies and the baselines mostly maintains the trend of Figure \ref{fig:rec_all}; that means higher NSR pertains to higher EXP and vice versa.

\begin{figure*}[htbp]
    \begin{subfigure}{0.24\textwidth}
    \includegraphics[width=\textwidth]{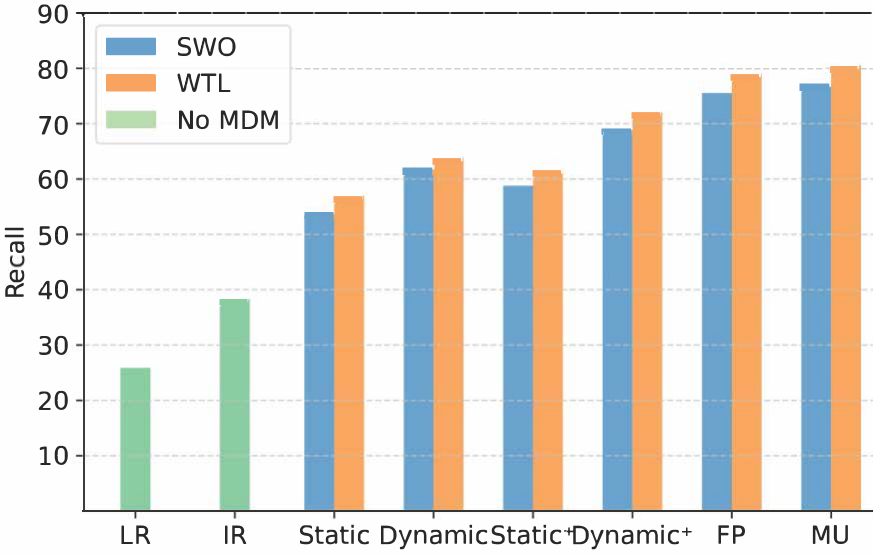}
    \caption{}
    \label{fig:ptb_mia_accross_dft_level}
    \end{subfigure}
    \hspace{0.5pt}
    \begin{subfigure}{0.24\textwidth}
    \includegraphics[width=\textwidth]{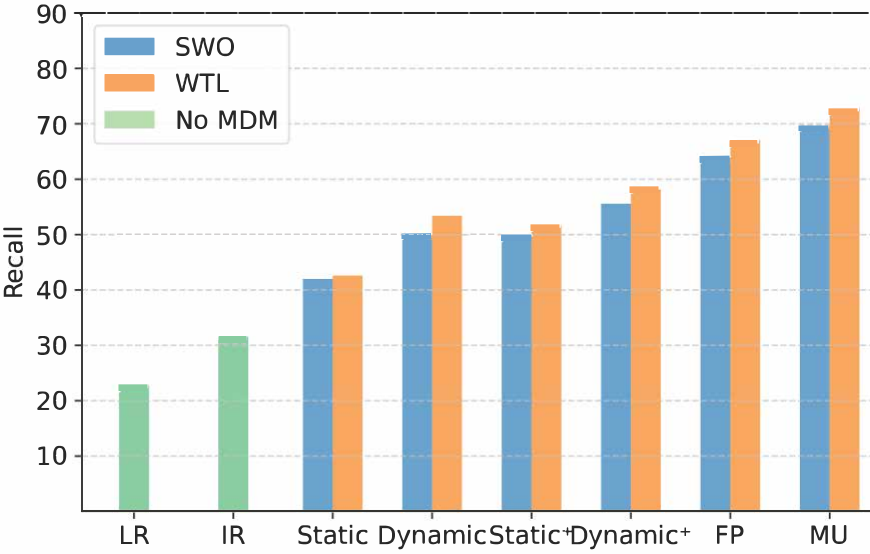}
    \caption{}
\label{fig:enron_mia_accross_dft_level}
    \end{subfigure}%
    \hspace{1pt}
    \begin{subfigure}{0.24\textwidth}
    \includegraphics[width=\textwidth]{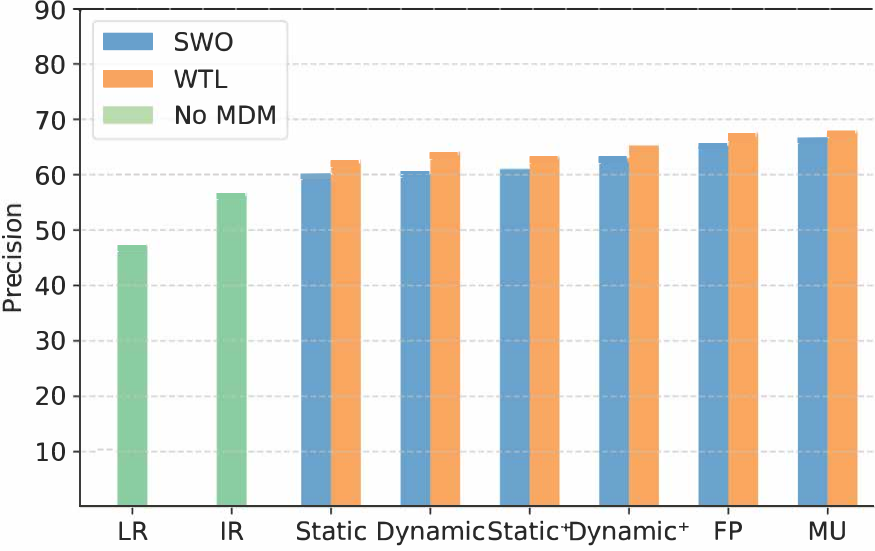}
    \caption{}
    \label{fig:ptb_mlm_precision}
    \end{subfigure}
    \hspace{0.5pt}
    \begin{subfigure}{0.24\textwidth}
    \includegraphics[width=\textwidth]{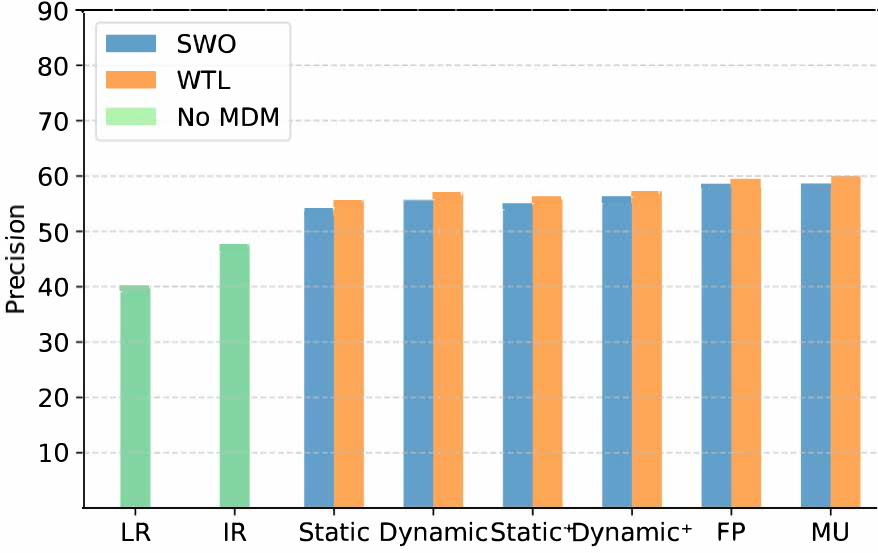}
    \caption{}
\label{fig:enron_mlm_precision}
    \end{subfigure}%
    
 \caption{Membership inference Recall and precision scores using BERT for different attack strategies along with three baselines. (a) and (c) are for PTB, (b) and (d) are Enron}
 \label{mia_all}
\end{figure*}

\begin{figure*}[h!]
\centering
    \begin{subfigure}{0.28\textwidth}
    \includegraphics[width=0.96\textwidth]{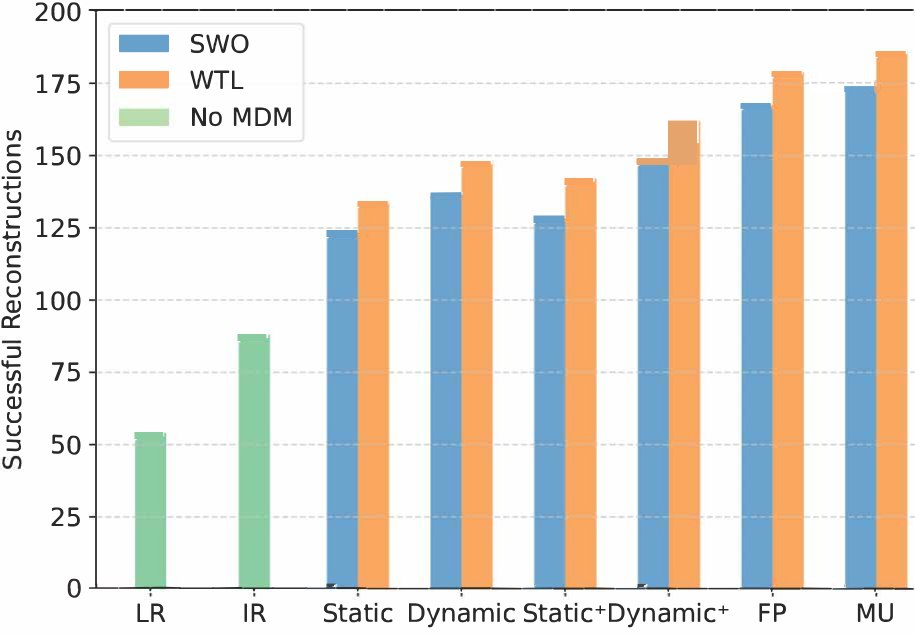}
    \caption{}
    \label{fig:ptb_Gemma_rec_accross_dft_level}
    \end{subfigure}
    \begin{subfigure}{0.27\textwidth}
    \includegraphics[width=0.95\textwidth]{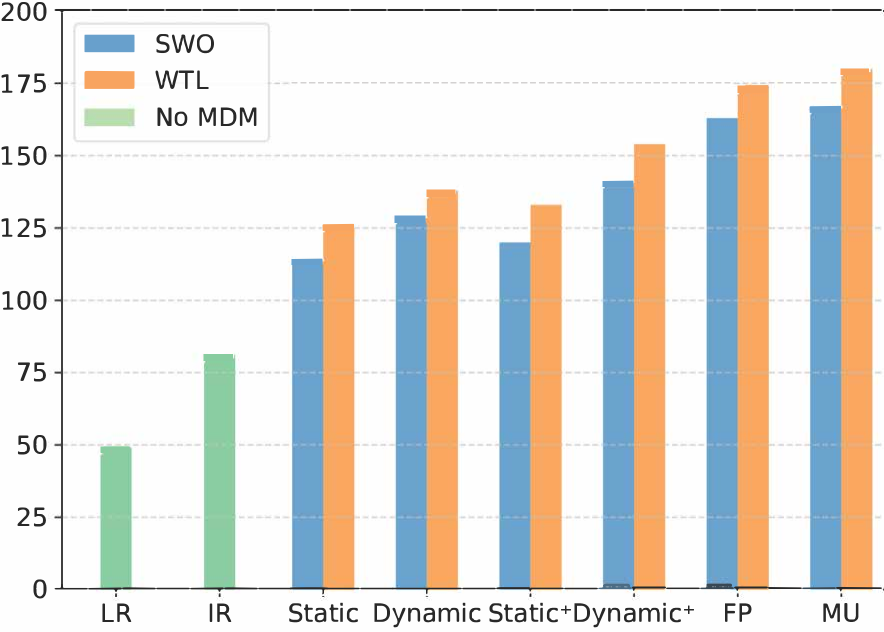}
    \caption{}
    \label{fig:ptb_rec_accross_dft_level}
    \end{subfigure}    
    \begin{subfigure}{0.27\textwidth}
    \includegraphics[width=0.96\textwidth]{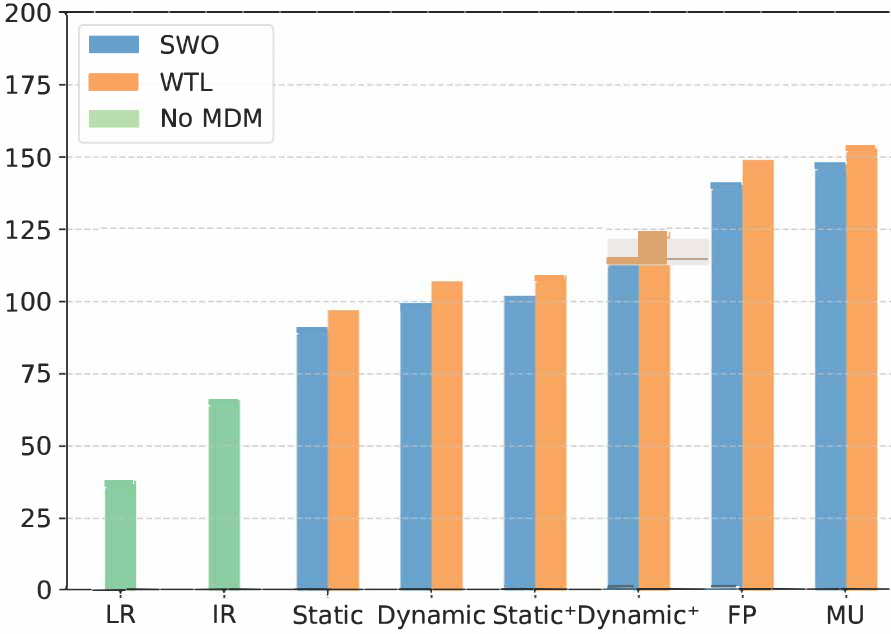}
    \caption{}
    \label{fig:ptb_mlm_rec_accross_dft_level}
    \end{subfigure}
    
 \caption{Number of successful reconstructions out of 200 unique canaries for different attack strategies along with the baselines on PTB dataset with (a)  Gemma, (b) GPT-2, and (c) BERT}
 \label{fig:ptb_all}
\end{figure*}



\begin{figure}[t]
\centering
    \begin{subfigure}{0.12\textwidth}
    \includegraphics[width=\textwidth]{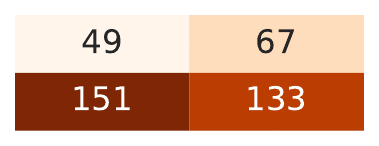}
    \caption{}
    \end{subfigure}%
    \begin{subfigure}{0.12\textwidth}
    \includegraphics[width=\textwidth]{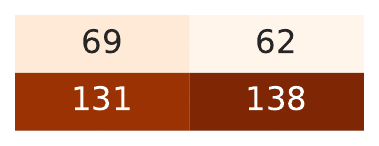}
    \caption{}
    \end{subfigure}
    \hspace{-5pt}
    \begin{subfigure}{0.12\textwidth}
    \includegraphics[width=\textwidth]{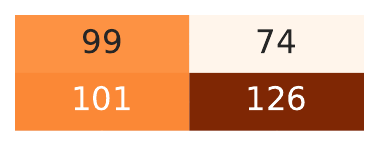}
    \caption{}
    \end{subfigure}%
    \begin{subfigure}{0.12\textwidth}
    \includegraphics[width=\textwidth]{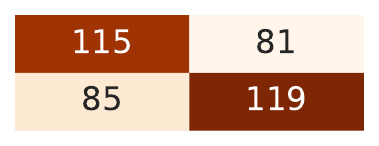}
    \caption{}
    \end{subfigure}
    
    \begin{subfigure}{0.12\textwidth}
    \includegraphics[width=\textwidth]{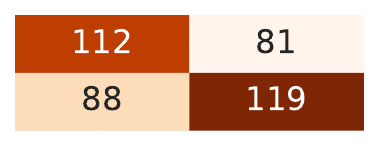}
    \caption{}
    \end{subfigure}%
    \begin{subfigure}{0.12\textwidth}
    \includegraphics[width=\textwidth]{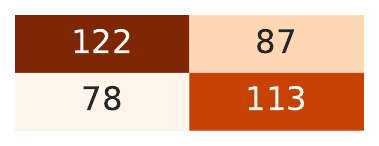}
    \caption{}
    \end{subfigure}
    \hspace{-5pt}
    \begin{subfigure}{0.12\textwidth}
    \includegraphics[width=\textwidth]{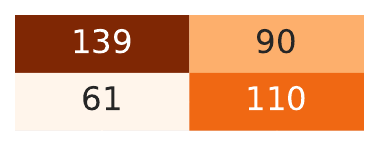}
    \caption{}
    \end{subfigure}%
    \begin{subfigure}{0.12\textwidth}
    \includegraphics[width=\textwidth]{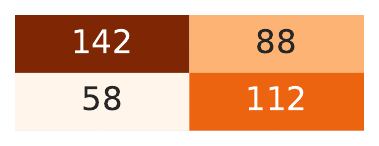}
    \caption{}
    \end{subfigure}
 \caption{Confusion Matrix of the Membership inference results on Wikitext dataset for (a) \BI~(b) \BII~(c) \offline~(d) \realtime~(e) \offlineXserver~(f) \realtimeXserver~(g) \BIII~(h) \BIV} 
 \label{fig:conf_mat_all}
\end{figure}

\section{Experiments with Llama-2}
\label{sec:llama-2}
As mentioned previously, we could not conduct experiments using Llama-2 in FL settings due to constraints on computation resources. Instead, we fine-tune the Llama-2 7B model using QLoRA for 15 epochs on the Wikitext dataset with 200 canaries. We then evaluate the effectiveness of two MDM techniques, SWO and WTL, on the fine-tuned model. In the trivial data reconstruction attack scenario, i.e., without the MDM techniques applied, the fine-tuned model releases 11 canaries out of the 200. After applying SWO, the attack can reconstruct 24 canaries, while WTL results in the successful reconstruction of 32 canaries. The NSR scores are significantly lower compared to the experiments with Gemma and GPT-2 (Figure \ref{fig:rec_all}) in the FL setup because the model is fine-tuned for a small number of epochs, giving it less opportunity to memorize those out-of-distribution canaries. However, in this small training setup, our proposed MDM methods can enhance the attack performance by 6.5\% (SWO) to 10.5\% (WTL) over the trivial attack. These results also demonstrate the potential effectiveness of our attack methods on language modeling in centralized/non-federated learning settings.
\section{Experiment with PTB Dataset}
\label{sec:ptb}
Results for the PTB dataset are shown in Figures \ref{fig:ptb_all}. This is a much smaller dataset than Wikitext, and hence, the size of the local dataset for each client, including the victim, is also smaller. In general, smaller datasets are more prone to overfitting than larger datasets \cite{halevy2009unreasonable, goodfellow2016deep}. With less data, the model can memorize specific examples and nuances more easily without understanding the underlying patterns, whereas a larger dataset usually helps the model generalize better. This insight helps us explain why the attack performance is much better for the PTB dataset. For instance, as shown in Figure \ref{fig:ptb_Gemma_rec_accross_dft_level}, the best performing \realtimeXserver~attack jumps to a successful reconstruction rate of 81\% (NSR=162). 
We can see similar growth in the attack performance with BERT in Figure~\ref{fig:ptb_mlm_rec_accross_dft_level}. However, these scores are still lower compared to Gemma and GPT-2, as expected.

\section{Membership Inference}
\label{{mi_appendix}}
\subsection{Implementation}
\label{MI_impl}
To infer membership of a sample $x$, i.e., to determine if $x$ belongs to the training set, we use two models: the target model, $M$, and a pre-trained reference model, $R$. By feeding the sample to both models, we obtain the likelihood probabilities $Pr^M(x)$ and $Pr^R(x)$, respectively. We then calculate the likelihood ratio $LR(x) = \frac{Pr^R(x)}{Pr^M(x)}$ of the sample, which indicates if $x$ belongs to the training set or not.  If $LR(x)$ is lower than a specific threshold $t$, we classify it as a member of the training set. Otherwise, we classify it as a non-member.
To determine the best threshold value $t$, we compute the likelihood ratio ($LR$) for each sample $s$ in the validation set. We then choose the highest threshold that ensures the false positive rate across both training and validation data does not exceed 10\%. Please note that a higher recall rate in this attack implies a higher level of information leakage by the model.

\subsection{Results for PTB and Enron}
\label{mia_ptb_enron}
In Figure \ref{mia_all}, we've displayed the membership inference results for PTB and Enron datasets. Looking at Figure \ref{fig:ptb_mia_accross_dft_level} and \ref{fig:enron_mia_accross_dft_level}, we observe that the \offline~strike has led to a significant improvement in the recall score. Specifically, for the PTB dataset, the recall score has increased by \textbf{14-27\%} compared to the lower baselines (\BI, \BII). Similarly, for the Enron dataset, we see a \textbf{10-19\%} improvement in the recall score when compared to the same baselines. As expected, the attack performance remained stable across different attack levels in terms of precision score for both datasets, as shown in Figure \ref{fig:ptb_mlm_precision} and \ref{fig:enron_mlm_precision}. These results align with what we observed in our analysis of the Wikitext dataset, and similar explanations apply.

\subsection{Confusion Matrix}
\label{subsec:confusiom_matrix}
In Figure \ref{fig:conf_mat_all}, we show the confusion matrix of the Membership inference results on Wikitext dataset. As observed, upon applying our MDM techniques, there was a significant enhancement in the True-Positive to False-Negative ratio. For instance, the static strike has raised the count of True Positives from \textbf{69} to \textbf{99} while reducing the count of False Negatives from \textbf{131} to \textbf{101} compared to the IR baseline. Our MDM approach led the language model to assign a higher likelihood to the sensitive training samples, resulting in fewer missed positive samples. Consequently, this resulted in a noticeable improvement in recall. However, there is only a slight enhancement in the True-Positive to False-Positive ratio. Therefore, the improvement in Precision is modest.

\section{Benchmark Study on FILM}
\label{bench_film}
The FILM attack \cite{gupta2022recovering} aims to recover private text data from information transmitted during training in federated learning.
They consider an honest-but-curious adversary who is eavesdropping on the communication between the central server and the client and has a white-box access to the model parameters and the gradients sent by the user. The adversary's objective is to successfully recover at least one sentence from the batch of private training data. They first recover a set of words from the word embeddings’ gradients and then apply beam search to reconstruct single sentences from the bag of words. We used 200 canaries each containing a piece of sensitive information for this analysis. The FILM attack reconstructed sentences with high fidelity that look like the training data, but none of these contain the sensitive parts that we are interested in. Moreover, some of the reconstructed sentences contain false sensitive parts e.g. a valid phone number that was not present in the canaries. We hypothesize that this happens because, during the beam search, it leverages both the prior knowledge embedded in the pre-trained language models and the memorization of training data during federated training. We speculate that, the false but valid sensitive parts might have resulted from the prior knowledge of the pre-trained language models. Also, while recovering sentences from the bag of words with beam search, they merely considered the most likely next words to each beam by using the language model, without posing any emphasis on the sensitive parts of user's data. Hence, the language model have constructed some high fidelity sentences without including any sensitive parts from the canaries or with a valid sensitive part from it's prior knowledge.

\section{Ablation Studies}
\label{sec:ab_app}
The following evaluations are done on GPT-2 base with the Wikitext dataset unless otherwise stated.

\subsection{Data Repetitions}
\label{data_rep}
Previous studies \cite{lee-etal-2022-deduplicating, carlini2021extracting} have shown that repeating sequences in the training set can increase memorization in language models. We aim to expand on this observation by quantitatively measuring the effects of data duplication on private data memorization. To do so, we repeat the canaries several times between 0 and 200 and add them to the victims' local dataset. Our results in Figure \ref{fig:result_with_repetition} demonstrate a clear log-linear trend in memorization. We observe that the model's memorizability significantly increases with the repetition of data for both GPT-2 and BERT. This is because the model tends to overfit on repeated data samples \cite{ying2019overview}.  
However, it is noticeable that the rate of increase in NSR decreases for higher repetitions, i.e., both curves get less steeper. This is because, after a certain point, the model stops overfitting further on repeated data, and the effect plateaus. 

\subsection{Prompt Perturbation}
\label{prompt_perturb}
To test the strength of our attack, we relax the assumption that the attacker has all of the private data prefixes in their original form. Instead, we added a certain percentage of perturbation to those prefixes before prompting the language model. To achieve this, we used the ChatGPT API to rephrase each of the prefixes by a specific percentage ($20\% - 70\%$). Next, we used the BERT-embedding-based similarity score \cite{kusner2015word} to determine the semantic coherence between each rephrased prefix and its original counterpart. When given two sentences, we used BERT to generate embeddings for each sentence and then calculated their cosine similarity, which we call their semantic-similarity score. The results of this analysis are shown in Figure \ref{fig:result_with_prompt_perturbation}, where we display the number of successful reconstructions against different ranges of semantic-similarity scores. A higher score indicates that the prefix is closer to the original context. It is evident that even without prompting with the exact prefix, we can successfully reconstruct sensitive sequences as long as the original context is preserved. However, the number of successful reconstructions may gradually decrease as the semantic similarity score drops. 

\subsection{Number of Sensitive Layers}
\begin{figure*}[t]
\centering
    \begin{subfigure}{0.3\textwidth}
    \includegraphics[width=\textwidth]{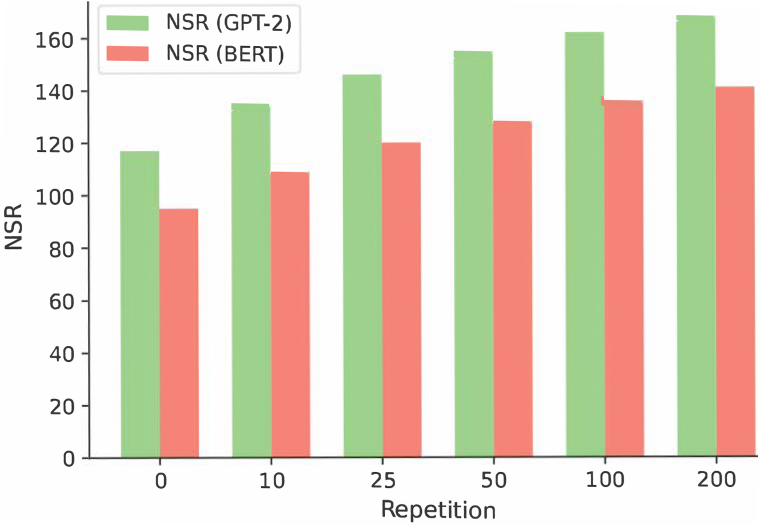}
    \caption{}
    \label{fig:result_with_repetition}
    \end{subfigure}%
    \hspace{5pt}
    \begin{subfigure}{0.36\textwidth}
    \includegraphics[width=\textwidth]{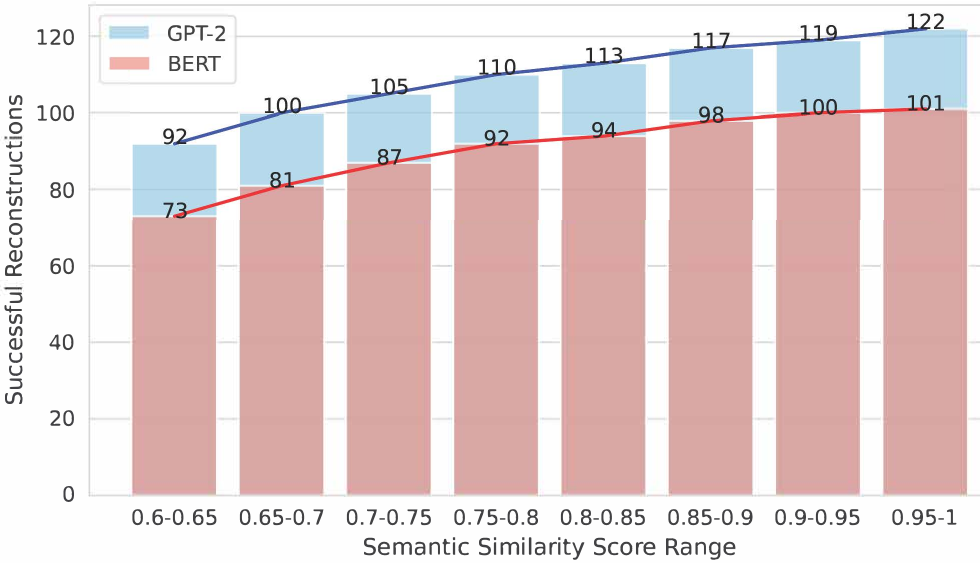}
    \caption{}
\label{fig:result_with_prompt_perturbation}
    \end{subfigure}
    \hspace{5pt}
    \begin{subfigure}{0.29\textwidth}
    \includegraphics[width=\textwidth]{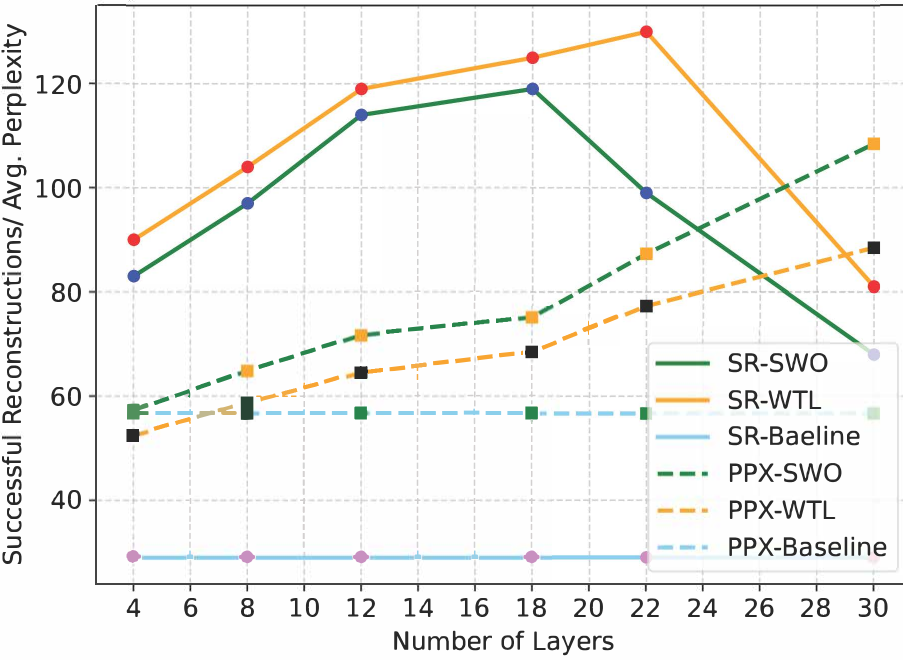}
    \caption{}
\label{fig:result_with_sens_layers}
    \end{subfigure}

 \caption{Results from ablation studies: leakage variation with (a) number of repetitions of the canaries  (b) number of prompt perturbations; lower semantic similarity refers to higher perturbation, (c)  number of layers in $L_{\star}$} 
 \label{fig:ablation_all}
\end{figure*}

Our proposed leakage boosting step involves fixing the size of $L_{\star}$, which refers to the number of sensitive layers to consider for weight adjustment. Figure \ref{fig:result_with_sens_layers} provides valuable insights into how this design choice affects attack performance and the main task performance. We conduct experiments for the \realtimeXserver attack by varying the number of sensitive layers from 4 to 30. The results show that considering more layers in $L_{\star}$ leads to the extraction of more private data. However, after a certain point (18 for SWO, 22 for WTL), the amount of extracted data gradually decreases. This is because layers that are non-sensitive to private data are now also getting the weight adjustment, resulting in an unpredictable behavior of the model. Due to the same reason,  increasing the size of $L_{\star}$ also causes a dip in the main task performance, as indicated by the model's perplexity on the attacker's local data (Figure \ref{fig:result_with_sens_layers}). 

\begin{figure*}[t]
\centering
    \begin{subfigure}{0.295\textwidth}
    \includegraphics[width=\textwidth]{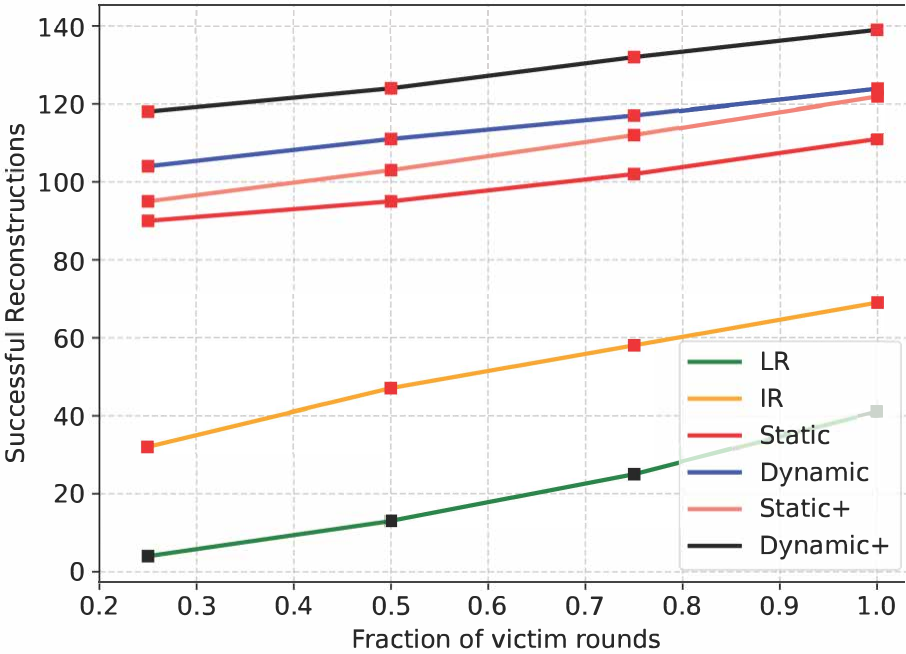}
    \caption{}
\label{fig:rec_with_vic_perc}
    \end{subfigure}
    \hspace{5pt}
    \begin{subfigure}{0.295\textwidth}
    \includegraphics[width=\textwidth]{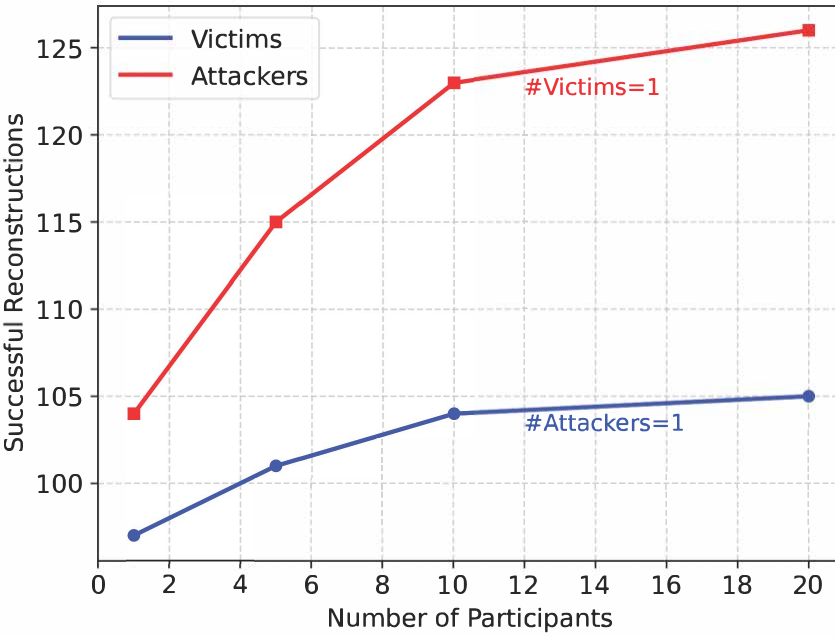}
    \caption{}
    \label{fig:rec_wth_att_vic}
    \end{subfigure}%
    \hspace{10pt}
    \begin{subfigure}{0.295\textwidth}
    \includegraphics[width=\textwidth]{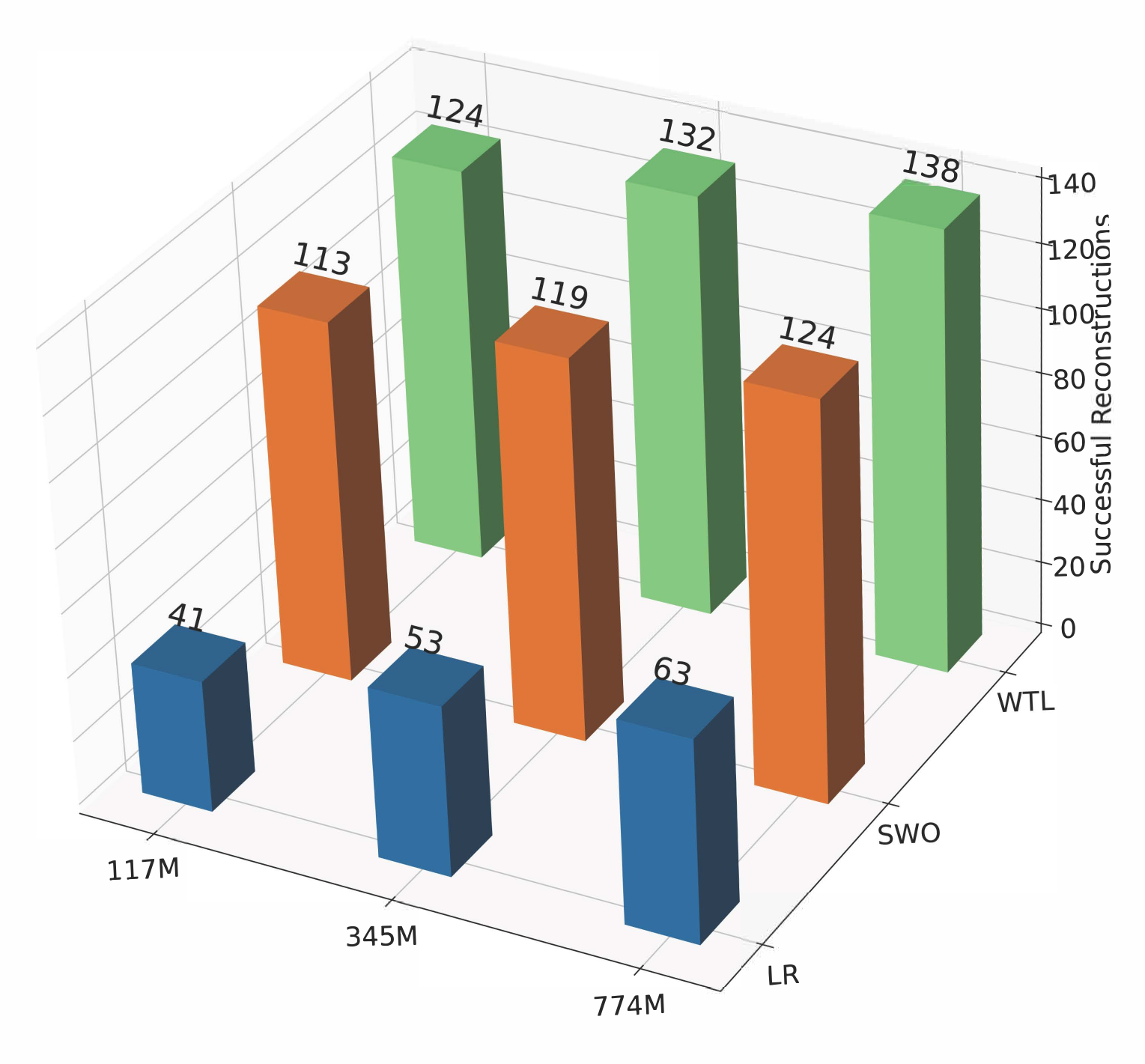}
    \caption{}
    \label{fig:rec_accross_model_sizes}
    \end{subfigure}

 \caption{Results from ablation studies: leakage variation with (a)  percentage of victim participation rounds. (b) number of victims and attackers (c) model sizes.} 
 \label{fig:ablation_app}
\end{figure*}
\subsection{Fraction of Victim rounds}
We intend to examine how the number of victim rounds affects the attack performance. Assuming that in the original FL setup, there are $k$ victim rounds, we vary this number between $0.25k$ and $k$. As shown in Figure \ref{fig:rec_with_vic_perc}, both baselines experience a rapid drop in NSR with a smaller proportion of victim rounds. Meanwhile, our four attack methods continue to maintain a high NSR. This indicates that our proposed strategies effectively reduce the impact of victim participation on privacy leakage.

\subsection{Number of Victims and Attackers:}
Figure \ref{fig:rec_wth_att_vic} depicts the variation in results due to changes in the number of victims and attackers, respectively. Note that the number of attackers has no impact during the \offline~mode of the attack; hence, relevant results are derived for the \realtime~attack. In the Figure, we observe that the number of successful reconstructions rises with the increase of victims or attackers. However, the rise is steeper with the increasing number of attackers because such an increase has a forthright influence during the \realtime~attack as the victim updates are getting pushed into the global model more frequently than before, leading to greater data leakage. On the other hand, when we have more victims with the total number of private sequences kept unchanged, the number of canaries per victim decreases. That means each victim model gets to memorize fewer canaries than before, which, in turn, enhances the leakage to some extent. But at the same time, increasing the number of victims hurts victim-round identification performance by increasing the number of false positive outcomes, as shown in Table \ref{tab:vic_idn}. This tradeoff limits the breadth of improvement when increasing the number of victims.
\subsection{Model Size:}
Prior work \cite{carlini2023quantifying} has shown that larger LMs memorize more than the smaller ones. 
Figure \ref{fig:rec_accross_model_sizes} shows a similar outcome in our experiment with three different sizes of GPT-2 model, as the NSR rises with the increasing model size. However, here, the noticeable event is that the increase in NSR for the MDM methods (SWO and WTL) is not as steep as \BI. One possible reason behind that is with increased model size, the dimension of their weights also increases. Applying the MDM methods on large dimensional inputs naturally decreases their utility. Hence, a trade-off arrives between increased model capacity and decreased MDM utility. Nevertheless, the NSR scores with the MDM methods are significantly higher than the baseline (\BI) for any model size.




\end{document}